\documentstyle[psfig,floats,preprint,aps]{revtex}
\tightenlines

\def\Section#1{}

\def\Sv{\vec{S}}
\def\Sz{S^z}
\def\Jp{{J'}}
\def\Jpc{{J'_c}}
\def\abs#1{\vert #1 \vert}
\def\Gap{{\sf E}}
\def\beq{\begin{equation}}
\def\eeq{\end{equation}}
\def\bea{\begin{eqnarray}}
\def\eea{\end{eqnarray}}
\def\nn{\nonumber}
\def\ie{{\it i.e.}}

\def\Reff{R_{{\rm eff.}}}
\def\age{\,\raise2pt\hbox{$\mathop{>}\limits_{\raise 2pt
\hbox{$\sim$}}$}\,}
\def\ale{\,\raise2pt\hbox{$\mathop{<}\limits_{\raise 2pt
\hbox{$\sim$}}$}\,}
\setbox122 = \hbox{1}
\def\id{\rlap{1}\rlap{\kern 1pt \vbox{\hrule width 4pt depth 0 pt}}
        \rlap{\kern 4 pt \hbox{\vrule height \ht122 depth 0 pt}}
	   \hskip\wd122}
\font\amsmath=msbm10 scaled \magstep1

\def\Zed{\hbox{\amsmath Z}}

\font\lasy=lasy10
\font\cmsy=cmsy10
\chardef\SymbolAchar="32
\chardef\SymbolBchar="33
\chardef\SymbolCchar="34
\def\SymbolA{\rlap{\kern0.8mm\raise0.3mm\hbox{$
           \cdot$}}\hbox{\lasy\SymbolAchar}}
\def\SymbolB{\rlap{\kern0.9mm\raise0.3mm\hbox{$
           \cdot$}}\hbox{\lasy\SymbolBchar}}
\def\SymbolC{\rlap{\kern1.0mm\raise0.1mm\hbox{$
           \cdot$}}\hbox{\cmsy\SymbolCchar}}
\def\href#1#2{{#2}}

\let\rdraft=\draft

\catcode`\@=11
\def\marginnote#1{}
\newcount\hour
\newcount\minute
\newtoks\amorpm
\hour=\time\divide\hour by60
\minute=\time{\multiply\hour by60 \global\advance\minute by-\hour}
\edef\standardtime{{\ifnum\hour<12 \global\amorpm={am}%
        \else\global\amorpm={pm}\advance\hour by-12 \fi
        \ifnum\hour=0 \hour=12 \fi
        \number\hour:\ifnum\minute<10 0\fi\number\minute\the\amorpm}}
\edef\militarytime{\number\hour:\ifnum\minute<10 0\fi\number\minute}
\def\draftlabel#1{{\@bsphack\if@filesw {\let\thepage\relax
   \xdef\@gtempa{\write\@auxout{\string
      \newlabel{#1}{{\@currentlabel}{\thepage}}}}}\@gtempa
   \if@nobreak \ifvmode\nobreak\fi\fi\fi\@esphack}
        \gdef\@eqnlabel{#1}}
\def\@eqnlabel{}
\def\@vacuum{}
\def\draftmarginnote#1{\marginpar{\raggedright\scriptsize\tt#1}}
\def\draft{\oddsidemargin -.5truein
        \def\@oddfoot{\sl preliminary draft \hfil
        \rm\thepage\hfil\sl\today\quad\militarytime}
        \let\@evenfoot\@oddfoot \overfullrule 3pt
        \let\label=\draftlabel
        \let\marginnote=\draftmarginnote
   \def\@eqnnum{(\theequation)\rlap{\kern\marginparsep\tt\@eqnlabel}%
\global\let\@eqnlabel\@vacuum}  }
\def\numberbysection{\@addtoreset{equation}{section}
        \def\theequation{\arabic{section}.\arabic{equation}}}
\def\underline#1{\relax\ifmmode\@@underline#1\else
        $\@@underline{\hbox{#1}}$\relax\fi}
\catcode`@=12
\relax
\def\ie{\hbox{\it i.e.}}        
        
\def\beq{\begin{equation}}
\def\eeq{\end{equation}}
\def\bea{\begin{eqnarray}}
\def\eea{\end{eqnarray}}
\def\nn{\nonumber}

\relax
\rdraft

\numberbysection

\begin{document}

\preprint{
\begin{minipage}[t]{2in}
\rightline{cond-mat/9802035}
\rightline{La Plata-Th 98/02}
\rightline{SISSA 14/98/EP}
\rightline{}
\end{minipage}
}

\title{Magnetization Plateaux in $N$-Leg Spin Ladders}

\author{D.C.\ Cabra$^{1}$, A.\ Honecker$^{2,\dag}$, P.\ Pujol$^{2}$}
\address{
$^{1}$Departamento de F\'{\i}sica, Universidad Nacional de la Plata,
      C.C.\  67, (1900) La Plata, Argentina {\rm and} \\
     Facultad de Ingenier\'{\i}a, Universidad Nacional de Lomas de Zamora, \\
     Cno.\ de Cintura y Juan XXIII, (1832), Lomas de Zamora, Argentina. \\
$^{2}$International School for Advanced Studies,
      Via Beirut 2-4, 34014 Trieste, Italy.\\
$^{\dag}$Work done under support of the EC TMR Programme
         {\em Integrability, non-per\-turba\-tive effects and symmetry in
         Quantum Field Theories}, grant FMRX-CT96-0012.\\
}

\date{February 3, 1998; revised April 8, 1998}
\maketitle
\begin{abstract}
\begin{center}
\parbox{14cm}{In this paper we continue and extend a systematic
study of plateaux in magnetization curves of antiferromagnetic
Heisenberg spin-$1/2$ ladders. We first review a bosonic
field-theoretical formulation of a single XXZ-chain
in the presence of a magnetic field, which is then used for an
Abelian bosonization analysis of $N$ weakly coupled chains. Predictions
for the universality classes of the phase transitions at the plateaux
boundaries are obtained in addition to a quantization condition for
the value of the magnetization on a plateau. These results are
complemented by and checked against strong-coupling expansions. 
Finally, we analyze the strong-coupling effective Hamiltonian
for an odd number $N$ of cylindrically coupled chains numerically.
For $N = 3$ we explicitly observe a spin-gap with a massive
spinon-type fundamental excitation and obtain indications that
this gap probably survives the limit $N \to \infty$.
}
\end{center}
\end{abstract}

\vspace{1cm}

\newpage

\section{Introduction}

So-called ``spin ladders'' have recently attracted a considerable
amount of attention (for reviews see e.g.\ \cite{review}).
They consist of coupled one-dimensional chains and may be
regarded as interpolating truly one- and two-dimensional systems.
Such an intermediate situation may be useful (among others) for
the understanding of high-$T_c$ superconductors. In fact,
modifications of the high-$T_c$ materials (see e.g.\ \cite{SrCuO})
give rise to experimental realizations of spin ladders. However,
the field was motivated by an observation that mainly concerns
the magnetic spin degrees of freedom, namely the appearance
of a spin gap in $N=2$ coupled gapless chains (see e.g.\
\cite{DRSBDRS}).

One-dimensional quantum magnets have been studied in great detail
over the past decades. One remarkable observation in this area
is the so-called ``Haldane conjecture'' \cite{Haldane} which states
that isotropic half-integer spin Heisenberg chains are gapless while
those with integer spin are gapped. Although this statement has
not been proven rigorously yet, a wealth of evidence supporting
this conjecture has accumulated in the meantime \cite{horda}
(see also \cite{CRP} for a recent field theoretical treatment).

Spin ladders are more general quasi one-dimensional quantum magnets.
Again, one of the attractions is a natural generalization of Haldane's
conjecture \cite{Haldane} to such $N$ coupled spin-$S$ chains:
If $S N$ is an integer, one expects a gap in zero field, otherwise not.
This conjecture is suggested among others by the large-$S$ limit (see
\cite{Sierra} for a recent review and references therein),
strong-coupling considerations \cite{DRSBDRS,RRT}, numerical computations
\cite{WNS,num,BGW} and even experiments (see e.g.\ \cite{SrCuO})
\footnote{This result was obtained for planar configurations and
often even an equal strength of coupling constants is assumed.
Such properties may be crucial as we will show below.
}.
If one includes a strong magnetic field, these Haldane gaps become
just a special case of plateaux in magnetization curves.
In the presence of a magnetic field, one of the central issues is the
quantization condition on the magnetization $\langle M \rangle$ for the
appearance of such plateaux, which for the purposes of the present
paper will be some special form of
\beq
l S N (1 - \langle M \rangle) \in \Zed \, .
\label{condMgen}
\eeq
Here $\langle M \rangle$ is normalized to a saturation value
$\langle M \rangle = \pm 1$ and $l$ is the number of lattice sites
to which translational symmetry is either spontaneously or explicitly
broken. Haldane's original conjecture \cite{Haldane} is related to
$l=N = 1$, $\langle M \rangle = 0$. More general cases for $N=1$ were
treated in \cite{AOY,Totsuka,Totsuka2}. In an earlier paper \cite{CHP}, we
have studied realizations of the condition (\ref{condMgen}) for
$N > 1$ but with the specializations $l=1$ and $S = 1/2$.

So far, spin ladders in strong magnetic fields have attracted
surprisingly little attention: To our knowledge, only the case of
an $N=2$ leg ladder had been investigated prior to
\cite{CHP}. The experimental measurement of the magnetization
curve of the organic two-leg ladder material
Cu$_2$(C$_5$H$_{12}$N$_2$)$_2$Cl$_4$ \cite{CCLMMP} gave rise
to theoretical studies using numerical diagonalization \cite{HLP},
series expansions \cite{WOS} and a bosonic field theory approach
\cite{ChiGi}.
In this case, (\ie\ $N=2$ and $S=1/2$) there is a spin gap which
gives rise to an $\langle M \rangle = 0$ plateau in the magnetization
curve. The transition between this zero magnetization plateaux
and saturation is smooth and no non-trivial effects (in particular
no symmetry breaking) were observed.

For $N>2$ one can expect plateaux at non-trivial $N$-dependent
fractions of the saturation magnetization \cite{CHP}. Though this
was a new observation for spin ladders, the phenomenon itself is not
completely new. For example, the possibility of magnetization
plateaux for the case of single spin-$S$ Heisenberg chains has been
discussed systematically in \cite{AOY} which also motivated some
of our work. The attraction of this phenomenon in spin ladders is
that they provide clear and natural realizations of such plateaux.
For example a numerical analysis of
the case $N=3$ explicitly exhibits a robust plateau with
$\langle M \rangle = 1/3$ \cite{CHP} which should also be
observable experimentally, e.g.\ in a suitable organic
spin-ladder material.

It is the purpose of the present paper to continue a systematic study
of (\ref{condMgen}) for generic $N\ge 1$ by using different complementary
techniques, such as Abelian bosonization, strong coupling expansions, and
numerical computations (the reader may find e.g.\ \cite{CHP}
helpful in the understanding of the present work). Here, we will
among others provide evidence that for $S = 1/2$ spontaneous breaking
of the translational symmetry to $l=2$ can be induced by strong
frustration or an Ising-like anisotropy, while $l \ge 3$ presumably
needs explicit symmetry breaking. (In \cite{Totsuka2} a slightly
different example of spontaneous symmetry breaking with $l=2$ was
studied).

The prefactor $l S N$ in (\ref{condMgen}) may seem quite cumbersome,
but it just counts the possible $S^z$-values
in a unit cell of a one-dimensional translationally invariant
groundstate. It should also be noted that (\ref{condMgen}) is just
a necessary condition; whether a plateau actually appears or
not depends on the parameters and the details of the model
under consideration. For example, plateaux with non-zero
$\langle M \rangle \ne 0$ have not been observed in the
$SU(2)$ symmetric higher spin-$S$ Heisenberg chains
(see e.g.\ \cite{BoPa}), unless translational invariance
is explicitly broken (c.f.\ \cite{Totsuka} for $S=1$).

Conditions of the type (\ref{condMgen}) occur also in generalizations
of the Lieb-Schultz-Mattis theorem \cite{LSM,Affleck,Rojo,AOY}.
This theorem constructs a {\it non-magnetic} excitation which
in the thermodynamic limit
is degenerate with the groundstate for a given magnetization
$\langle M \rangle$ and orthogonal to it unless $\langle M \rangle$
satisfies (\ref{condMgen}). In this manner one proves the existence
of either gapless excitations or spontaneous breaking of translational
symmetry. Unfortunately it is at present not clear that this
theorem applies to plateaux in magnetization curves since they
require a gap to {\it magnetic} excitations.

Here we concentrate mainly on the case $S=1/2$ and all couplings in the
antiferromagnetic regimes, but try to keep $N$ as general as possible.
Other situations can be analyzed as well, but may lead to somewhat
different physics (compare e.g.\ \cite{HKSprep} for the example
of $N=3$ antiferromagnetically coupled ferromagnetic chains).

This paper is organized as follows: In section II we first review
some aspects of the
formulation of a single XXZ-chain as a bosonic $c=1$ conformal
field theory. This serves as a basis of later investigations
and illustrates some generic features also present in
$N$-leg spin ladders. In section III we first introduce the
precise lattice model and its field-theoretic counterpart
which we then analyze in the weak-coupling regime. Section IV
starts from the other extreme --the strong-coupling limit--
and proceeds with series expansions around this limit. In section V we
discuss an effective Hamiltonian for the strong-coupling limit
of an odd number $N$ of cylindrically coupled chains which
we then analyze numerically in section VI. We summarize
our results by presenting ``magnetic phase diagrams'' in
section VII before we conclude with some comments and open problems
(section VIII).

\section{A single XXZ-chain}

First, we recall some results for the XXZ-chain on a ring of
$L$ sites in the presence of a magnetic field $h$ applied along the $z$-axis:
\beq
H_{XXZ} = J \sum_{x=1}^L \left\{ \Delta \Sz_x \Sz_{x+1} + {1 \over 2}
\left(S^{+}_x S^{-}_{x+1} + S^{-}_x S^{+}_{x+1} \right)\right\}
- h \sum_{x=1}^L \Sz_{x} \, .
\label{HamXXZ}
\eeq
Apart from being the basis for the investigation in later sections,
this also serves as an illustration of some general features.
It should be noted that in (\ref{HamXXZ})
the magnetic field is coupled to a conserved quantity which is
related to the magnetization $\langle M \rangle$ via
$\langle M \rangle = \left\langle {2 \over L} \sum_{x=1}^L \Sz_{x}
\right\rangle$. For this reason, properties of (\ref{HamXXZ})
in the presence of a magnetic field $h \ne 0$ can be related to
those at $h = 0$ and the magnetic field $h$ can be considered
as a chemical potential.

The Hamiltonian (\ref{HamXXZ}) is exactly solvable by Bethe ansatz
also for $h \ne 0$. In this way it can be rigorously shown that
its low-energy properties are described by a $c=1$ conformal field
theory of a free bosonic field compactified at radius $R$ in
the thermodynamic limit for $\Delta > -1$ and any given magnetization
$\langle M \rangle$ (see e.g.\ \cite{HaldXXZ} and compare also
\cite{WET} for a detailed discussion of the case $\abs{\Delta} < 1$).
More precisely, upon insertion of the bosonized representation of the spin
operators into the Hamiltonian (\ref{HamXXZ}) (see e.g.\ \cite{AffLe})
one obtains the following low-energy effective Hamiltonian
for the XXZ-chain:
\beq
\bar{H}_{XXZ} = \int {\rm d}x {\pi \over 2} \left\{ \Pi^2(x) +
    R^2(\langle M \rangle, \Delta) \left(\partial_x \phi(x)\right)^2
\right\}
\label{bosHamXXZ}
\eeq
with $\Pi = {1 \over \pi} \partial_x \tilde{\phi}$, and
$\phi = \phi_L + \phi_R$, $\tilde{\phi} = \phi_L - \phi_R$.
In (\ref{bosHamXXZ}) we have suppressed a for our purposes
irrelevant proportionality constant that includes the
velocity of sound. In this formulation, the effect of both
the magnetic field $h$ and XXZ-anisotropy $\Delta$ turns
up only via the radius of compactification
$R(\langle M \rangle, \Delta)$. This radius governs the
conformal dimensions, in particular the conformal dimension
of a vertex operator ${\rm e}^{i \beta \phi}$ is given by
$\left({\beta \over 4 \pi R}\right)^2$. We now describe how $R$ 
can be computed.

We parametrize the XXZ-anisotropy by $\Delta = \cos\theta$ with
$0 < \theta < \pi$ for $-1 < \Delta < 1$ and by $\Delta = \cosh\gamma$
with $\gamma > 0$ for $\Delta > 1$. Now for given magnetization
$\langle M \rangle \ge 0$ and XXZ-anisotropy $\Delta$, the associated
radius of compactification $R$ and magnetic field $h/J$ can be
obtained by solving integral equations (see e.g.\
\cite{WET,Bogo,KBIbook,QFYOA}) in the following way:
Firstly, introduce a function $\sigma(\eta)$ for the density of
particles satisfying the integral equation
\beq
\sigma(\eta) = {1 \over 2 \pi} \left\{ g(\eta) - \int_{-\Lambda}^{\Lambda}
         K(\eta-\eta') \sigma(\eta') {\rm d}\eta'\right\}
\label{intEqSi}
\eeq
where the kernel $K(\eta)$ and the righthand side $g(\eta)$ are presented
in Table \ref{funTab}.

\begin{table}[ht]
\begin{tabular}{c|ccc}
$\Delta$ & $K(\eta)$ & $g(\eta)$ & $\epsilon_0(\eta)$ \\ \tableline
$\cos\theta = \Delta < 1$ & 
    ${\tan\theta \over \tan^2\theta \; \cosh^2{\eta \over 2}
                  +\sinh^2{\eta \over 2}}$ &
      ${\cot{\theta\over 2} \over \cosh^2{\eta \over 2}
                  + \cot^2{\theta\over 2} \; \sinh^2{\eta \over 2}}$ &
         ${h \over J} - {\sin^2{\theta} \over \cosh{\eta} - \cos{\theta}}$ \\
$\Delta =1$ &
    ${4 \over \eta^2 + 4}$ &
      ${2 \over \eta^2 + 1}$ &
         ${h \over J} - {2 \over \eta^2 + 1}$ \\
$\cosh\gamma = \Delta > 1$ &
    ${\tanh\gamma \over \tanh^2\gamma \; \cos^2{\eta \over 2}
                  +\sin^2{\eta \over 2}}$ &
      ${\coth{\gamma\over 2} \over \cos^2{\eta \over 2}
                  + \coth^2{\gamma\over 2} \; \sin^2{\eta \over 2}}$ &
         ${h \over J} - {\sinh^2{\gamma} \over \cos{\eta} - \cosh{\gamma}}$ \\
\end{tabular}
\caption{Functions appearing in the integral equations.}
\label{funTab}
\end{table}

The real parameter $\Lambda \ge 0$ in (\ref{intEqSi}) describes
the spectral parameter value at the Fermi surface and is determined
by the magnetization $\langle M \rangle$ via the filling condition
\beq
\int_{-\Lambda}^{\Lambda} \sigma(\eta) {\rm d}\eta
= {1 \over 2} \left(1 - \langle M \rangle\right) \, .
\label{detmLam}
\eeq
In general, one has to adjust $\Lambda$ iteratively by first
numerically solving (\ref{intEqSi}) and then checking for (\ref{detmLam}).
Only some special cases can be solved explicitly. This includes
the case $\langle M \rangle = 0$ and $\Delta \le 1$ where $\Lambda = \infty$
is the correct choice. Once the desired value of $\Lambda$ is
determined, one introduces a dressed charge function $\xi(\eta)$
(see e.g.\ \cite{Bogo,KBIbook}) as a solution of the integral equation
\beq
\xi(\eta) = 1 - {1 \over 2 \pi} \int_{-\Lambda}^{\Lambda}
         K(\eta-\eta') \xi(\eta') {\rm d}\eta'
\label{intEqXi}
\eeq
giving directly rise to the radius of compactification
\beq
R(\langle M \rangle, \Delta) = {1 \over \sqrt{4 \pi} \; \xi(\Lambda)} \, .
\label{rocIntEq}
\eeq
If one further wants to determine the associated magnetic field $h$,
one has to introduce another function $\epsilon_d(\eta)$, the dressed energy,
satisfying the integral equation
\beq
\epsilon_d(\eta) = \epsilon_0(\eta) - {1 \over 2 \pi} \int_{-\Lambda}^{\Lambda}
         K(\eta-\eta') \epsilon_d(\eta') {\rm d}\eta'
\label{intEqEpsd}
\eeq
with the bare energy $\epsilon_0(\eta)$ listed in Table \ref{funTab}.
Then the magnetic field $h/J$ is determined by the condition that
the energy of the dressed excitations vanishes at the Fermi
surface
\beq
\epsilon_d(\Lambda) = 0 \, .
\label{vanCondEps}
\eeq
Using (\ref{intEqXi}) one can see that $\epsilon(\eta) =
\epsilon(\eta)\vert_{h=0} + {h \over J} \xi(\eta)$ solves
(\ref{intEqEpsd}) if $\epsilon(\eta)\vert_{h=0}$ solves
(\ref{intEqEpsd}) with formally $h=0$ (but for the given
$\langle M \rangle$). From this and (\ref{vanCondEps}) one
can easily obtain $h$ once $\Lambda$ is known via
$h/J = -\epsilon(\Lambda)\vert_{h=0} / \xi(\Lambda)$.

In general, these integral equations have to be solved numerically
and $\Lambda$ has to be determined by some iterative method.
Although this is readily
done by standard methods, a generally accessible implementation
seems to be still unavailable. We have therefore decided to
tentatively provide access to such solutions on the WWW \cite{WWW}.
This implementation works in the way described above. Typically,
it gives results with an absolute accuracy
of $10^{-6}$ or better. Of course, one can change the order
of the procedure: For example one could also prescribe $h/J$,
then determine $\Lambda$ from (\ref{intEqEpsd}) and (\ref{vanCondEps}),
next the radius of compactification $R$ from (\ref{intEqXi}) and
(\ref{rocIntEq}) and finally (optionally) the magnetization
$\langle M \rangle$ from (\ref{intEqSi}) and (\ref{detmLam}).

Some remarks are in order concerning the case $\Delta > 1$: In this
region the identification with a $c=1$ conformal field theory has
been established directly only by a numerical analysis of the
Bethe-ansatz equations (for a summary see e.g.\ section 2 of \cite{AlMa}).
However, the six-vertex model in external fields has been shown
\cite{NoKi} to yield a $c=1$ conformal field theory, and since the
XXZ-chain arises as the Hamiltonian limit of the six-vertex model,
this indirectly establishes the identification. Still, explicit
formulae e.g.\ for $R$ are not available in the literature and
therefore we have obtained the integral equations presented
for the case $\Delta > 1$ above by an analytical continuation of
those for $\Delta < 1$, as is suggested e.g.\ by \cite{BdVV} (see
also \cite{ISbook} -- apparently such a continuation was also used
in the recent work \cite{Totsuka}). We have performed some checks that
this yields indeed correct results. For example, some radii $R$
associated to certain magnetic fields $h$ and anisotropies $\Delta$
obtained numerically for chains of length up to $L=234$ \cite{AlcPri}
are reproduced in this way.

As a further check, one can compare the critical magnetic field for the
boundary of the $\langle M \rangle = 0$ plateau at $\Delta > 1$
obtained by numerical solution of the above integral equations
with the exact solution of the Bethe-ansatz equations \cite{ClGa}
\beq
{h_c \over J} = {2 \pi \sinh{\gamma} \over \gamma} \sum_{n=0}^{\infty}
              {1 \over \cosh{{(2 n + 1) \pi^2 \over 2 \gamma}}} \, ,
\label{hcXXZ}
\eeq
where as before $\Delta = \cosh{\gamma}$. In this case (\ie\ for
$\langle M \rangle = 0$) one has $\Lambda = \pi$ and
the above integral equations can be solved using Fourier series
\cite{BdVV,ISbook,ClGa}. If one solves the above integral equations
numerically, the deviation from the exact result (\ref{hcXXZ}) is of
the order of the numerical accuracy
(in our implementation \cite{WWW} always less than $10^{-6}$).

In passing we make a comment which will turn out to be useful later. 
The result (\ref{hcXXZ}) is the gap to $\Sz = 1$ excitations.
However, the fundamental excitation of the XXZ-chain is known
to be a so-called ``spinon'' which carries $\Sz = 1/2$
\cite{FaTa,GoSa,RHSZ}. This spinon can be regarded as a domain-wall
between the two antiferromagnetic groundstates for $\Delta > 1$.
Since a single spin-flip creates two domain-walls,
the lowest $\Sz = 1$ excitation is a scattering state of two
spinons. This picture can be useful e.g.\ in numerical computations.
For example, a single spinon can be observed for odd $L$ with
periodic boundary conditions.

\begin{figure}[ht]
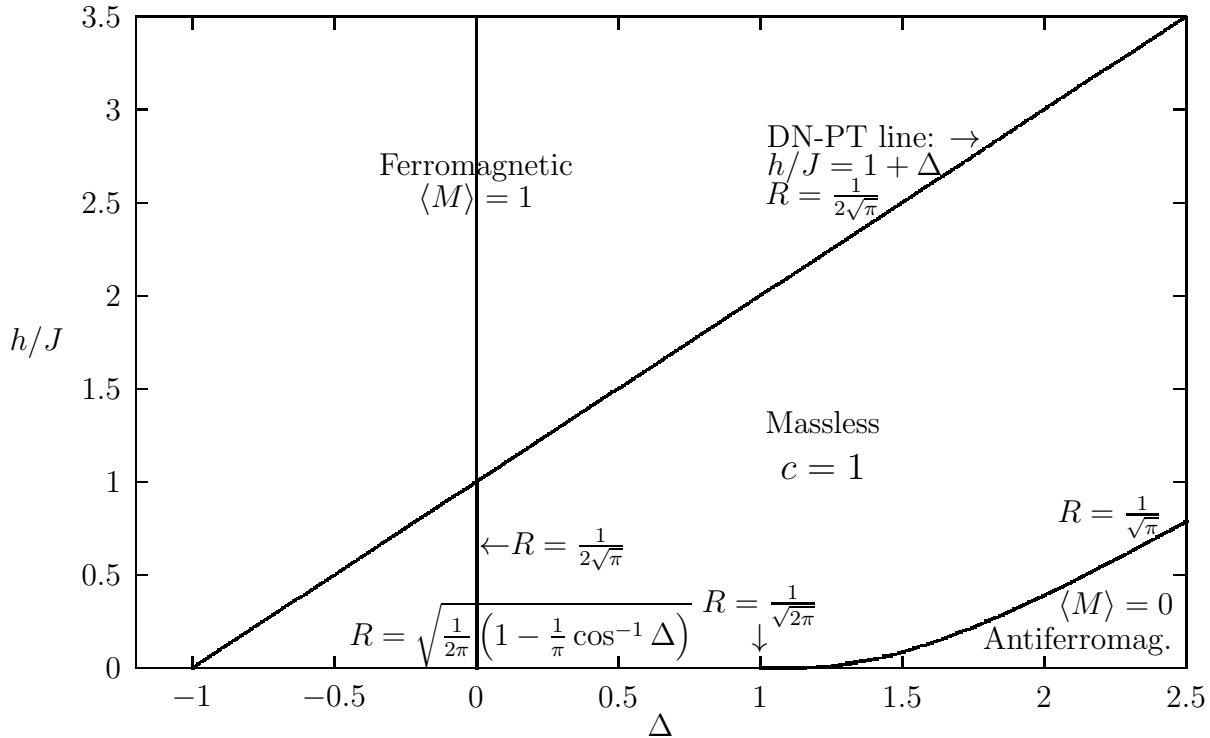

\input xxz-fig.tex
\caption{
Magnetic phase diagram of the XXZ-chain (\ref{HamXXZ}). For
explanations compare the text.
\label{xxzfig}
}
\end{figure}

After this digression let us now return to the above integral equations.
The results obtained from them are summarized in
the magnetic phase diagram for the XXZ-chain Fig.\ \ref{xxzfig}
(see also \cite{JoMcC,AlMa} for similar pictures). There are
two gapped phases: A ferromagnetic one at sufficiently strong
fields (which is actually the only one for $\Delta < -1$) and
an antiferromagnetic phase for $\Delta > 1$ at small fields.
In between is the massless phase where the bosonized form
(\ref{bosHamXXZ}) is valid. An elementary computation of the
spinwave dispersion above the ferromagnetic groundstate shows
that the transition between the ferromagnetic
phase and the massless phase is located at
$h/J = 1 + \Delta$. This transition is a very clear example
of the DN-PT universality class \cite{DzNe,PoTa} \footnote{
This terminology is motivated as follows: This type of transition
is usually known under the name ``Pokrovsky-Talapov''
\cite{PoTa}, but in the context of magnetization processes
was actually discovered first by Dzhaparidze and Nersesyan \cite{DzNe}.
}, \ie\ for $\langle M \rangle \to 1$ the magnetization 
behaves as (compare also \cite{HoPa})
\beq
\left(\langle M \rangle - M_c\right)^2 \sim h^2 - h_c^2
\label{DNPTexp}
\eeq
with here $M_c = 1$ and $h_c/J = 1 + \Delta$. At this transition
the radius takes the universal value $R(1,\Delta) = {1 \over 2 \sqrt\pi}$.

The other transition line starts at the $SU(2)$ symmetric point
$\langle M \rangle = 0$, $\Delta = 1$ with a radius
$R(0,1) = {1 \over \sqrt{2 \pi}}$. Actually, at $\langle M \rangle = 0$
the additional operator
\beq
\cos\left(4 \sqrt{\pi} \phi\right)
\label{XXZgapOp}
\eeq
appears in the continuum limit, which we have suppressed
in (\ref{bosHamXXZ}) since it is irrelevant inside the massless
phase. At $\Delta = 1$ it is marginal and becomes
relevant for $\Delta > 1$, opening the gap that gives the boundary
of the antiferromagnetic phase in Fig.\ \ref{xxzfig}. The associated
phase transition is a Kosterlitz-Thouless (K-T) transition \cite{KoTh}
(see e.g.\ \cite{Luther,AGG,Nijs}). The almost marginal nature of the
operator responsible for the gap leads to a stretched exponential
decay for $\Delta$ slightly bigger than one which is characteristic
for a K-T transition \cite{Kosterlitz}. The exact asymptotic form
for the gap (or critical magnetic field) of the XXZ-chain is
easily obtained from the Bethe-ansatz solution (\ref{hcXXZ})
upon noting that $\gamma \approx \sqrt{2 (\Delta - 1)}$ and that 
in the limit $\Delta \to 1$ only the term $n=0$ contributes to the
sum. One then finds
\cite{ClGa}:
\beq
\qquad\qquad
{h_c \over J} \sim 4 \pi {\rm e}^{-{\pi^2 \over 2 \sqrt{2 (\Delta-1)}}}
\qquad\qquad \hbox{(for $\Delta$ slightly bigger than $1$).}
\label{hcXXZcl1}
\eeq
For this reason the phase boundary is indistinguishable from
the $h = 0$ line for XXZ-anisotropies up to $\Delta
\approx 1.2$ on the scale of Fig.\ \ref{xxzfig}. In this region,
the numerical determination of the radius $R$ is difficult for
$\langle M \rangle \to 0$. Nevertheless, using that $\Lambda = \pi$
for $\langle M \rangle = 0$ and $\Delta > 1$, one can readily check
that the constant function $\xi(\eta) = {1 \over 2}$ solves the
integral equation (\ref{intEqXi}). Then one obtains from
(\ref{rocIntEq}) that $R(0,\Delta>1)= {1 \over \sqrt{\pi}}$.

For $\langle M \rangle = 0$ and $\abs{\Delta} \le 1$ one
has $\Lambda = \infty$ (see above) and one can use the Wiener-Hopf
method to solve the above integral equations in closed form.
This yields in particular $R(0,\Delta)=
\sqrt{{1 \over 2\pi}\left(1-{1 \over \pi}\cos^{-1}\Delta \right)}$
\cite{AffLe}. In general,
the radius $R$ increases with increasing $\Delta$. For $\Delta > 0$,
it decreases with increasing magnetization $\langle M \rangle$, while
for $\Delta < 0$ this is reversed to an increase in $R$ with
increasing $\langle M \rangle$. Clearly, the radius must be
constant on the XY-line which separates these two regions:
$R(\langle M \rangle, 0) = {1 \over 2 \sqrt\pi}$ (though the
magnetic field associated to a given $\langle M \rangle$ is
still a non-trivial function which should be computed from the
above integral equations). 

\begin{figure}[ht]
\psfig{figure=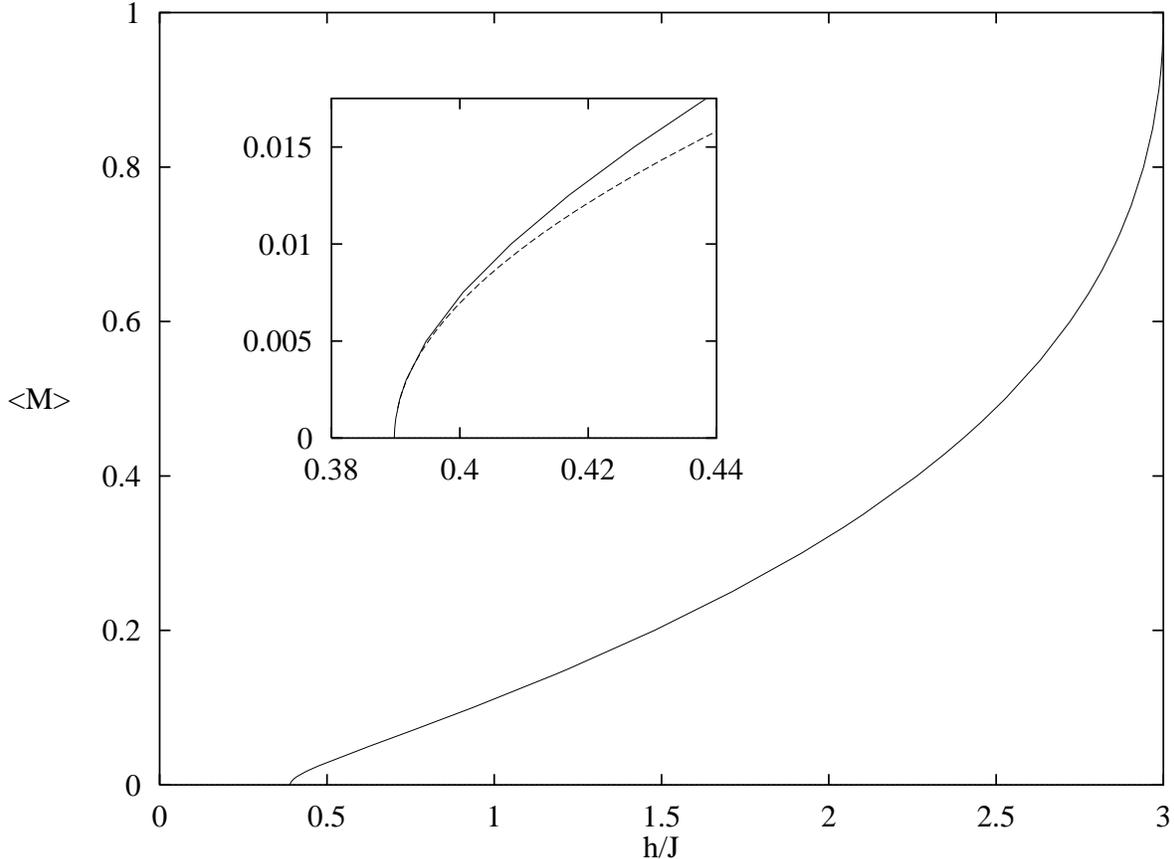,width=\columnwidth,angle=270}
\caption{
Magnetization curve of the XXZ-chain at $\Delta = 2$ obtained
from the integral equations (\ref{intEqSi}), (\ref{detmLam}),
(\ref{intEqEpsd}) and (\ref{vanCondEps}). The inset
shows the region of small magnetization and
illustrates that also the transition $\langle M \rangle \to 0$
is compatible with the DN-PT universality class for $\Delta > 1$
(the dashed line is a fit to the universal form (\ref{DNPTexp})).
\label{figMagD2}
}
\end{figure}

Including the operator (\ref{XXZgapOp}) in the bosonized language
(\ref{bosHamXXZ}), one recovers a Hamiltonian treated
in \cite{SchulzCI} as a model for commensurate-incommensurate
transitions. This means that the transition $\langle M \rangle \to 0$
for $\Delta > 1$ is predicted to be in the DN-PT universality class
\cite{DzNe,PoTa}, too. The same bosonization argument also leads to
the already mentioned result $R(0,\Delta>1) = {1 \over \sqrt{\pi}}$
(see also \cite{HaldXXZ}).
The inset in Fig.\ \ref{figMagD2} illustrates
for $\Delta = 2$ that for sufficiently small magnetizations
one can indeed observe a behaviour that is compatible with the
universal square-root (\ref{DNPTexp}). It should be noted though that
the window for the universal DN-PT behaviour is too small to permit
verification within our numerical accuracy for smaller values of $\Delta$
(e.g.\ $\Delta \le 1.2$) where neither a reliable numerical
check of the result $R= {1 \over \sqrt{\pi}}$ is possible.
An analytic check of the asymptotic behaviour of the magnetization
from the Bethe-ansatz solution would be interesting, but
is beyond the scope of the present paper.

It should be noted that the height of the entire inset in Fig.\
\ref{figMagD2} corresponds to the first step in
the magnetization curve of a chain of the finite size $L = 112$.
Therefore, the exact solution is crucial in verifying the
exponent (\ref{DNPTexp}) -- a numerical or experimental verification of
this behaviour restricted to such a small region would be extremely
difficult. Similar difficulties will be faced in an experimental or
numerical verification of $R(0,\Delta) = {1 \over \sqrt{\pi}}$ or
the equivalent statement for the correlation function exponents in
the region of $\Delta$ slightly larger than 1.

\section{Ladders: Weak coupling and Abelian bosonization}

In this section we apply Abelian bosonization to the weak-coupling
region $\Jp \ll J$ of $N$-leg spin ladders. In particular, we will
show how the necessary condition (\ref{condMgen}) arises in this
formulation and discuss under which circumstances an allowed plateau
does indeed open as a function of the parameters $\Jp$ and $\Delta$.
The lattice Hamiltonian for this system is given by
\beq
H^{(N)} = \Jp \sum_{i,j} \sum_{x=1}^L \Sv_{i,x} \Sv_{j,x}
     + J \sum_{i=1}^N \sum_{x=1}^L \left\{
\Delta \Sz_{i,x} \Sz_{i,x+1} + {1 \over 2}
\left(S^{+}_{i,x} S^{-}_{i,x+1} + S^{-}_{i,x} S^{+}_{i,x+1} \right)\right\}
     - h \sum_{i,x} \Sz_{i,x} \, ,
\label{hamOp}
\eeq
where $\Jp$ and $J$ are respectively the interchain and intrachain
couplings, $h$ is the external magnetic field and the indices $i$
and $j$ label the different chains (legs) in the ladder. The sum
in the first term is over all possible couplings between chains.
The case of periodic boundary conditions (PBC) and open boundary
conditions (OBC) will be discussed later. 
Here we have explicitly included an XXZ-anisotropy $\Delta$ in the
intrachain coupling. We have kept the interchain coupling $\Jp$
$SU(2)$ symmetric for simplicity in later sections although this
is not substantial in the weak-coupling region which we will
discuss in the remainder of this section.

The corresponding effective field-theoretic Hamiltonian is obtained
using standard methods \cite{SchBo,AffLe} (see also
\cite{AOY,Totsuka,Totsuka2} for the case of non-zero magnetization).
One essentially uses (\ref{bosHamXXZ}) as the effective Hamiltonian
for each chain and the bosonized expressions for the spin operators
which read:
\beq
\Sz_{i,x} \approx \frac{1}{\sqrt{2\pi}} \frac{\partial \phi_i}{\partial x}
 +const. : \cos(2k_F^i x+\sqrt{4\pi}\phi_i) :
 + {\langle M_i\rangle \over 2} \, ,
\label{7}
\eeq
and
\beq
S^{\pm}_{i,x}\approx \ : {\rm e}^{\pm i\sqrt{\pi} \tilde{\phi}_i}(1+const.
\cos(2k_F^i x+\sqrt{4\pi}\phi_i)) : \, .
\label{8}
\eeq
Here we have set a lattice constant to unity which appears in passing
to the continuum limit. The colons denote normal ordering which we
take with respect to the groundstate of a given mean magnetization
$\langle M_i \rangle$ in the $i$th chain which is a natural choice.
This leads to the constant term in (\ref{7}) which will play an
important r\^ole in the discussion of the terms that can be generated
radiatively. The prefactor $1/2$ arises from our normalization of
the magnetization to saturation values $\langle M \rangle = \pm 1$.
The Fermi momenta $k_F^i$ are given by
$k_F^i = \pi (1- \langle M_i \rangle)/2$.

In the weak-coupling limit along the rungs, $\Jp \ll J$,
we obtain the following bosonized low-energy effective Hamiltonian for
the $N$-leg ladder keeping only the most relevant perturbation terms:
\bea
\bar{H}^{(N)} &=& \int {\rm d}x \Biggl[
      {\pi \over 2} \sum_{i=1}^N \left\{ \Pi_i^2(x) +
    R^2(\langle M \rangle, \Delta) \left(\partial_x \phi_i(x)\right)^2
\right\}
\nn \\
& +& {\lambda_1 \over 2 \pi} \sum_{i,j} \left(\partial_x \phi_i(x)\right)
      \left(\partial_x \phi_{j}(x)\right) \nn \\
& +& \sum_{i,j} \biggl\{
            \lambda_2 : \cos(2 x (k_F^i + k_F^j)
             + \sqrt{4\pi}(\phi_i + \phi_{j})) :
\label{LeH}  \\
&&\!\!\! + \lambda_3 : \cos(2 x (k_F^i - k_F^j)
             + \sqrt{4\pi} (\phi_i - \phi_{j})) :
         + \lambda_4 : \cos\left(\!\sqrt{\pi} (\tilde{\phi}_i -
\tilde{\phi}_{j})
                             \! \right) \! :
    \biggr\}\Biggr] \, , \nn
\eea
The four coupling constants $\lambda_i$ essentially correspond
to the coupling $\Jp$ between the chains: $\lambda_i \sim \Jp/J$.
In arriving to the Hamiltonian (\ref{LeH}) we have discarded a constant term
and absorbed a term linear in the derivatives of the free bosons into a
redefinition of the applied magnetic field.

The Hamiltonian (\ref{LeH}) has been also used to represent spin-$N/2$
chains (see e.g.\ \cite{SchBo,Schulz}), since they can be obtained in
the limit of $N$ strongly ferromagnetically coupled chains
($\Jp \to -\infty$). However, here we will analyze (\ref{LeH})
mainly in the case of small antiferromagntic $\Jp$ and discuss
various boundary conditions.

Note that the $\lambda_2$ and $\lambda_3$ perturbation terms contain
an explicit dependence on the position (in the latter case this
$x$-dependence disappears for symmetric configurations with equal $k_F^i$).
Such operators survive in passing from the lattice to the
continuum model, assuming that the fields vary slowly, only when they
are commensurate. In particular, the $\lambda_2$ term appears in
the continuum limit only if the oscillating factor
$\exp(i2x (k_F^i + k_{F}^j))$ equals unity. If the configuration
is symmetric, this in turn happens only for zero magnetization
(apart from the trivial case of saturation).

For simplicity let us first analyze the case with $N=3$ 
and PBC. We first have to diagonalize the Gaussian (derivative) part of 
the Hamiltonian. This is achieved by the following change of variables 
in the fields:
\beq
\psi_1 = {1\over \sqrt{2}} \left( \phi_1 - \phi_3 \right)
\, , \qquad
\psi_2 = {1\over \sqrt{6}} \left( \phi_1 + \phi_3 - 2 \phi_2 \right)
\, , \qquad
\psi_D = {1\over \sqrt{3}} \left( \phi_1 + \phi_2 + \phi_3 \right)
\, .
\eeq
In terms of these fields the derivative part of the Hamiltonian can be 
written as:
\beq
\bar{H}_{{\rm der.}} = \int {\rm d}x 
      {\pi \over 2}  \left\{ 
    R^2(\langle M \rangle, \Delta) \left[
(1 + a) \left(\partial_x \psi_D(x)\right)^2  + 
(1 - b ) \left(\left(\partial_x \psi_1(x)\right)^2 + 
\left(\partial_x \psi_2(x)\right)^2 \right) \right]
\right\}
\label{der}
\eeq
where $a= 2\Jp/(J \pi^2 R^2) = 2 b$.
We can now study the large-scale behaviour of the effective Hamiltonian
(\ref{LeH}) where we assume all $k_F^i$ equal due to the symmetry of
the chosen configuration of couplings. Let us first consider the case
when the magnetization $\langle M \rangle$ is non-zero.
In this case only the $\lambda_3$ and $\lambda_4$ terms are present.
The one-loop R.G.\ equations are:
\bea
\frac{db}{d\ln{L}}&=&4\pi\left(-\frac{3\lambda_3^2}{2R^2}+
12\pi^2R^2\lambda_4^2\right)
\nonumber \\
\frac{d\lambda_3}{d\ln{L}}&=&\left( 2- \frac{1}{2\pi R^2 (1-b)} \right)\lambda_3
-\pi \lambda_3^2
\nonumber \\
\frac{d\lambda_4}{d\ln{L}}&=&\left( 2- 2\pi R^2 (1-b) \right)\lambda_4
-\pi \lambda_4^2 \, .
\label{firstrg}
\eea
It is important to notice that only the fields $\psi_1$ and $\psi_2$
enter in these R.G.\ equations, since the perturbing operators
do not contain the field $\psi_D$. The behaviour of these R.G.\
equations depends on the value of $R$. The main point is that always
one of the two $\lambda$ perturbation terms will dominate and the corresponding
cosine operator tends to order the associated fields. This gives a
finite correlation length in correlation functions containing the
fields $\psi_1$ and $\psi_2$ (or their duals). For example,
for $\Delta \le 1$ we have that $R^2 < (2 \pi)^{-1}$ since
$\langle M \rangle \neq 0$. Then, from (\ref{firstrg}) one can
easily see that the dominant term will be the $\lambda_4$ one.
This term orders the dual fields associated with $\psi_1$ and 
$\psi_2$. Then, the correlation functions involving these last fields
decay exponentially to zero. In either case, the field $\psi_D$ remains 
massless. A more careful analysis of the original Hamiltonian shows that
this diagonal field will be coupled to the massive ones only through very
irrelevant operators giving rise to a renormalization of its compactification
radius. However, due to the strong irrelevance of such coupling terms these
corrections to the radius are expected to be small, implying that the 
value of the large-scale effective radius keeps close to the zero-loop
result $R \sqrt{1-a}$. It is straightforward
to generalize this to $N$ chains when all possible coupling are present
and have the same value $\Jp$. 
One can find a change of variables on the fields to:
$$
\psi_D~~;~~ \psi_i~~~i=1, \ldots, N-1
$$
where $\psi_D = {1/\sqrt{N}} \sum_{i=1}^{N} \phi_i$. Again, for non-zero 
magnetization, all but the diagonal field $\psi_D$ will be present
in the perturbing terms $\lambda_3$ and $\lambda_4$. The R.G.\ equations
are essentially the same as (\ref{firstrg}) and the result is that only
the field $\psi_D$ will be massless.

We are then left in principle with a free Gaussian action
for the diagonal field. However some operators can be radiatively
generated. We see from eqs.~(\ref{7},\ref{8}) that when we turn on
the interchain coupling, the ``N-Umklapp'' term
\beq
\Jp^N \cos\left(2x\sum_{i=1}^{N} k_{F}^i
   + \sqrt {4\pi} \sum_{i=1}^{N}\phi_i\right)
=\Jp^N \cos\left(2x\sum_{i=1}^{N} k_{F}^i + \sqrt {4\pi N} \psi_D \right)
\label{tu}
\eeq
appears in the operator product expansion (OPE).

Again, this operator survives in passing from the lattice to the continuum
model, assuming that the fields vary slowly, only when the oscillating
factor $\exp(i2x\sum_{j=1}^{N} k_{F}^j)$ equals one. This in turn will happen
when the following specialized version of the condition (\ref{condMgen}) 
\beq
{N \over 2} (1 - \langle M \rangle) \in \Zed \, 
\label{condM}
\eeq
is satisfied.
At such values of the magnetization, the field $\psi_D$ can then undergo
a K-T transition to a massive phase, indicating the presence of a
plateau in the magnetization curve.
An estimate of the value of $\Jp$ at which this operator becomes
relevant can be obtained from its scaling dimension which in zero-loop
approximation is given by
\beq
\dim\left(\cos\left( \sqrt {4\pi N} \psi_D \right)\right) =
  {N \over 4\left(\pi R^2+{N-1 \over \pi} \; {\Jp \over J} \right)}
\, .
\label{dimPsiD}
\eeq
At $\Delta = 1$ one then obtains $\Jpc \approx 0.09 J$ for the
$\langle M\rangle = 1/3$ plateau at $N=3$ and
$\Jpc \approx 0.7 J$ for $\langle M \rangle = 1/2$ at $N=4$
and also for $\langle M \rangle = 1/5$ at $N=5$.
At the opening of such plateaux, the 
effective radius of compactification is fixed to be:
\beq
\Reff^2 = {N \over 8\pi}
\label{reff}
\eeq
and the large-scale effective spin operators are (c.f.\ \cite{SchBo}):
\beq
\Sz_{{\rm eff.}}(x)\approx
  \sqrt{{N \over 2\pi}} \frac{\partial \psi_D}{\partial x}
 +const. : \cos(2k_F x+\sqrt{4\pi N}\psi_D) :
+ {\langle M\rangle \over 2} \, ,
\label{zeff}
\eeq
and
\beq
S^{\pm}_{{\rm eff.}}(x)\approx \ :
   {\rm e}^{\pm i\sqrt{\pi/N} \tilde{\psi}_D}(1+const.
   \cos(2k_F x+\sqrt{4\pi N}\psi_D)) : \, .
\label{xyeff}
\eeq
Then, eq.\ (\ref{reff}) fixes the values of the correlation exponents 
at this point to be
\beq
\eta_{z} = 4  \, ; \qquad \eta_{xy} = {1 \over 4} \, .
\label{ExpKT}
\eeq
On the other hand, commensurate-incommensurate transition results 
\cite{SchulzCI,ChiGi,Totsuka} imply that the values of these exponents
should be
\beq
\eta_{z} = 2 \, ; \qquad \eta_{xy} = {1 \over 2}
\label{ExpBD}
\eeq
along the upper and lower
boundary of a plateau. This situation is similar to the XXZ-chain at
$\Delta = 1$ and $\Delta > 1$ for the boundary of the
$\langle M \rangle = 0$ plateau. However, the values of the
exponents are different since the perturbing operators are
different.

Note that the ``N-Umklapp" process which allows the appearance of (\ref{tu})
produces a complete family of operators given by:
\beq
\cos\left(2xl\sum_{i=1}^{N} k_{F}^i + l\sqrt{4\pi} \sum_{i=1}^{N}\phi_i\right)
= \cos\left(2xl\sum_{i=1}^{N} k_{F}^i +l \sqrt {4\pi N} \psi_D \right)
\label{tutu}
\eeq
with $l$ an arbitrary integer. The values of the magnetization for which
one of these operators is allowed are subject to a generalization
of (\ref{condM}), namely (\ref{condMgen}) in the Introduction (with
$S = 1/2$).
However, the dimensions of these operators increase with $l^2$. So,
these operators cannot be relevant unless we consider regimes with an
anisotropy parameter $\Delta$ bigger than one or very big values
of the interchain coupling $\Jp$ far from the perturbative regime of
the present analysis. Therefore, higher values of $l$ are realized
only under special conditions.
While $l=2$ can be obtained by either strong Ising-like anisotropy
$\Delta$ or frustration at strong coupling (see section VI below),
it is possible that $l \ge 3$ can be realized only if suitable
symmetry breaking terms are explicitly introduced into the Hamiltonian
(\ref{hamOp}).

Note that formally, the preceding analysis can also be carried out using
the fermionic Jordan-Wigner formulation. For example, in this
formulation the ``N-Umklapp'' operator (\ref{tu}) is given by
$$\left(\prod_{a=1}^N R_a^\dag(x) L_a(x) \exp(2 i k_F^a x)\right)
+ \left(\prod_{a=1}^N L_a^\dag(x) R_a(x) \exp(-2 i k_F^a x)\right) \, ,$$
where $R_a$ and $L_a$ are the right- and left-moving components of
the fermions. We have chosen to use the bosonized language
because it is more appropriate for general values of the anisotropy
$\Delta$.

The analysis above was for the case where all the chains were
coupled together with the same coupling value. More precisely, the 
estimates for the appearance of plateaux were for positive
(frustrating) interchain coupling. To generalize this to PBC
(which is different from the preceding case for $N \ge 4$),
we first notice, using the bosonized expression of the effective
Hamiltonian, that this configuration of couplings is not stable
under R.G.\ transformation. E.g.\ the OPE between terms like
$\cos(\phi_1-\phi_2)$ and $\cos(\phi_2-\phi_3)$ generates
an effective coupling between the fields $\phi_1$ and $\phi_3$, etc.
The underlying intuitive picture is that antiferromagnetic couplings
between the chain $2$ with the chains $1$ and $3$ generates
an effective ferromagnetic coupling between the chains $1$ and $3$.
For example, for $N=4$ and PBC, ferromagnetic couplings are generated
along the diagonals between originally uncoupled chains. This case
is part of the family of configurations with antiferromagnetic
nearest neighbour and ferromagnetic next-nearest neighbour couplings. 
For this general situation at $N=4$,
the coupling matrix in the derivative part is given by: 
\beq
\pmatrix{1 & a & -b& a \cr
         a & 1 & a & -b \cr
        -b & a & 1 & a \cr
         a & -b& a & 1 \cr
}
\eeq
where $a$ and $b$ are positive.
As in the preceding analysis, one can change variables to
\bea
&\displaystyle
 \psi_D = {1 \over \sqrt{4}}(\phi_1+\phi_2+\phi_3+\phi_4) \, ; \quad
&\psi_1 = {1 \over \sqrt{4}}(\phi_1-\phi_2+\phi_3-\phi_4) \, ; \nn \\
&\displaystyle
 \psi_2 = {1 \over \sqrt{2}}(\phi_1-\phi_3) \, ;
&\psi_3 = {1 \over \sqrt{2}}(\phi_2-\phi_4) \, .
\eea
For generic values of the magnetization,
it is easy to see that the diagonal field $\psi_D$ is again
the only field that does not acquire a mass under the perturbation.
Then, the analysis of the appearance of the N-Umklapp term for
particular values of the magnetization is identical to the one performed
before. The generalization to generic $N$ with PBC is straightforward,
one first builds the radiatively generated couplings by keeping only
the lowest order in $\Jp$. Once this step is performed, the only 
difference with respect to the case of equal interchain couplings
is the zero-loop value of the dimension of the N-Umklapp operator
(which enters via the initial conditions for the R.G.\ flow). This has
the effect of changing the value of the coupling $\Jp$ at which a plateau
opens with a given value of the magnetization, but the qualitative behaviour
of the system is similar. This conclusion is not so straightforward
for $\langle M \rangle =0$, where as we will see, the difference between
frustrating and non-frustrating configurations can become crucial.

Concerning finally the case of OBC, let us first consider again the case
$N=3$ with antiferromagnetic coupling between the first and second chain
and the second and the third chain. Again, this coupling is not stable
under R.G.\ transformation. 
Under R.G.\ transformations the OBC configuration will flow
towards a non-frustrating cyclically coupled configuration.
The main point is that for weak coupling and non-zero magnetization,
the most relevant perturbing term will be again the one containing
differences of fields or their duals. Then they will produce a mass gap
for all the relative degrees of freedom and one recovers a scenario
similar to the symmetric case, where only one massless field 
is left. On the other hand, the appearance of ``N-Umklapp" operators
and their commensurability is unchanged, since these criteria depend
on the value of the magnetization and not on the particular couplings
between the chains.

Let us study now the more complicated case of zero magnetization.
For $\langle M \rangle = 0$ 
the $\lambda_2$ term in (\ref{LeH}) is commensurate and must be included in
the perturbation terms. The situation is now much more complicated because this
relevant operator couples the diagonal field $\psi_D$ with the massive ones.
For equal coupling between $N=3$ chains, the R.G.\ equations are now:
\bea
\frac{da}{d\ln{L}}&=&\frac{16\pi \lambda_2^2}{R^2}
\nonumber \\
\frac{db}{d\ln{L}}&=&4\pi\left(-\frac{\lambda_2^2}{2R^2}-\frac{3\lambda_3^2}{2R^2}+
12\pi^2R^2\lambda_4^2\right)
\nonumber \\
\frac{d\lambda_2}{d\ln{L}}&=&\left(2-\frac{2}{3}
\frac{1}{4\pi R^2}\left(\frac{2}{1+a} +\frac{1}{1-b}\right) \right)\lambda_2
- \pi \lambda_2 \lambda_3
\nonumber \\
\frac{d\lambda_3}{d\ln{L}}&=&\left( 2- \frac{1}{2\pi R^2 (1-b)} \right)\lambda_3
- \pi \lambda_3^2 - \pi \lambda_2^2
\nonumber \\
\frac{d\lambda_4}{d\ln{L}}&=&\left( 2- 2\pi R^2 (1-b) \right)\lambda_4
- \pi \lambda_4^2
\label{rg2}
\eea
with the R.G.\ initial conditions
\beq
a(0) = 2b(0) = {2\Jp \over J \pi^2 R^2},
\label{incon}
\eeq
and
\beq
\lambda_2(0)=\lambda_3(0)=1/2 (const)^2 \Jp/J~;~\lambda_4(0)=\Jp/J 
\label{inlamb}
\eeq
where we kept the notation of (\ref{LeH},\ref{der}). We see that the 
radius of compactification of the diagonal field is now strongly affected
by the presence of the $\lambda_2$ term. 
Note also that the ``N-Umklapp" process generates the operator
\beq
\Jp^N\cos\left(\sqrt{4\pi N} \psi_D \right)
\label{even}
\eeq
for $N$ even, and
\beq
\Jp^N\cos\left(2\sqrt{4\pi N} \psi_D \right)
\label{odd}
\eeq
for $N$ odd.
For non-frustrating interchain coupling 
(a negative $\Jp$ coupling between all the chains for example),
all relative fields are massive according to \cite{Schulz}.
We can then integrate out these massive degrees of freedom. The crucial 
point is that now the radius of the diagonal field gets a non-trivial
correction due to the strong interaction with the massive fields. Since
this field is the only one expected to describe the large-scale behaviour
of the system, for $\Delta = 1$ and
$\langle M \rangle=0$, the $SU(2)$ symmetry of the model would fix
the radius of this field to be \cite{Schulz}:
\beq
\Reff = \sqrt{N \over {2\pi}} \, .
\eeq
For such a value of the renormalized radius, the ``N-Umklapp" term becomes
strongly relevant for $N$ even, and marginally irrelevant for $N$ odd. 
These arguments are based on the assumption that the (uncontrolled) R.G.\
flow will drive our system to the ($SU(2)$ symmetric) strong coupling 
regime. The situation is even more subtle for positive $\Jp$ 
(or $\lambda_i$), 
because in this case, from (\ref{rg2}) one sees that the quadratic terms
could now prevent the R.G.\ flow to reach the same strong coupling regime
as for $\Jp < 0$. A numerical analysis of the R.G.\ flow for a frustrated
three-leg Hubbard ladder at half filling provides evidence for the
opening of a gap \cite{Arrigoni}. On the other hand, a non-abelian
bosonization analysis \cite{ToSu} lead to the conclusion that the
weak-coupling region is gapless.
This case deserves further investigation and series
expansions are one way to approach this issue.

\section{Strong coupling expansions for $N$-leg ladders}

In this section we diagonalize the interaction along the rungs
exactly for $J = 0$ and then expand quantities of interest in
powers of $J/\Jp$ around this limit. In order to be able to
cover a variety of cases, we used a quite general method to
perform the series expansions which is summarized e.g.\ in 
section 3 of \cite{Pert} (actually, the program used in the
present paper is a modified version of the one used loc.cit.).

As was already pointed out in \cite{CHP}, one can simply count
the number of chains $N$ in the limit $J/\Jp \to 0$ in order
to determine the allowed values of the magnetization $\langle M \rangle$.
This is presumably the simplest way to obtain the quantization condition
(\ref{condM}). A less trivial fact is that all these values of the
magnetization are in fact realized. For example, for ferromagnetic
coupling $\Jp < 0$, the magnetization jumps immediately from
one saturated value $\langle M \rangle = -1$ to the other one
($\langle M \rangle = +1$) as the magnetic field $h$ passes
through zero. Nevertheless, for not too large $N$ one can readily
compute the magnetization curve of (\ref{HamXXZ}) and check for
antiferromagnetic coupling $\Jp > 0$ that all possible values
of the magnetization are indeed successively realized as the field
is increased.
The critical magnetic fields $h$ at which one value of
the magnetization jumps to the next largest one are
given in Table \ref{strongCoupTab}.

\begin{table}[ht]
\begin{tabular}{lcc}
\multispan3 $h/\Jp$ \\ \tableline
$N$ & {\it OBC} & {\it PBC} \\ \tableline
$2$ & $\pm 1$ & \hbox to 1.5 cm{\hrulefill} \\
$3$ & $\pm {3 \over 2}$, $0$ & $\pm {3 \over 2}$, $0$ \\
$4$ & $\pm \left(1 + {1 \over \sqrt{2}}\right)$,
           $\pm {1 + \sqrt{3} - \sqrt{2} \over 2}$ &
            $\pm 2$, $\pm 1$ \\
$5$ & $\pm {5 + \sqrt{5} \over 4} = \pm 1.80902$, $\pm 1.11887$, $0$ &
            $\pm {5 + \sqrt{5} \over 4}$, $\pm {3  + \sqrt{5} \over 4}$, $0$
                        \\
$6$ & $\pm 1.86603$, $\pm 1.38597$, $\pm 0.49158$ &
            $\pm 2$, $\pm {1 + \sqrt{5} \over 2}$,
           $\pm {\sqrt{13} - \sqrt{5} \over 2}$ \\
\end{tabular}
\caption{Values of $h$ at which the magnetization jumps for (\ref{HamXXZ})
with coupling constant $\Jp$, $\Delta = 1$, $N$ sites and different
boundary conditions.}
\label{strongCoupTab}
\end{table}

As a next step, one can take the intrachain coupling $J$ perturbatively
into account.
First, we look at a two-leg ladder ($N=2$).
The rung Hamiltonian $H_r = \Jp \Sv_1 \Sv_2$ has two eigenvalues
whose difference corresponds to the critical fields presented in
Table \ref{strongCoupTab}.
The lower eigenvalue equals $-3 \Jp / 4$ and belongs to the
spin $S=0$ eigenstate while the other threefold degenerate one equals
$\Jp / 4$ and corresponds to the spin triplet ($S=1$).
For convenience, we concentrate on an isotropic interaction
for the rungs, but it is straightforward to include an XXZ-anisotropy
$\Delta$ in the interaction along the chains. One motivation
for doing so is that this permits further comparison with the
weak-coupling analysis ($\Jp \ll J$) of the previous section.
At $J = 0$ the groundstate is obtained by putting singlets on each rung.
A basic excitation at $J = 0$ is given by one triplet in a sea
of singlets. Since the $SU(2)$ symmetry is broken down to $U(1)$
by the perturbation, different series are obtained for the
$S^z = \pm 1$ and $S^z = 0$ components of the triplet.

Here we concentrate just on the series for the gap, but also previous
results for the groundstate energy and the dispersion relations
are readily extended to higher orders \cite{RRT}, or to analytical
expressions in $\Delta$ for longer series \cite{WKO} at $\Delta = 1$
with numerical coefficients.

The gap is obtained by the value of the excitation energy of a
single flipped spin at momentum $k=\pi$ with $S^z = \pm 1$.
We find
\bea
{\Gap_2 \over \Jp} &=& 1 - \left({J \over \Jp}\right)
   + {1 + \Delta^2 \over 4} \left({J \over \Jp}\right)^2
   + {(1 + \Delta)^2 \over 16} \left({J \over \Jp}\right)^3
   + {-2 + 6 \Delta - 9 \Delta^2 + \Delta^4 \over 32}
                        \left({J \over \Jp}\right)^4 \nn \\
&& + {21 - 84 \Delta + 39 \Delta^2  - 48 \Delta^3  + 2 \Delta^4 \over 256}
                        \left({J \over \Jp}\right)^5 \nn \\
&& - {82 - 98 \Delta + 155 \Delta^2  - 50 \Delta^3  + 80 \Delta^4  - 12 \Delta^6
                       \over 1024} \left({J \over \Jp}\right)^6
   + {\cal O}\left(\left({J \over \Jp}\right)^7\right) \, .
\label{GapSerN2}
\eea
At the isotropic point $\Delta = 1$ we recover well-known
results: For this special case, the first three
orders can be found in \cite{RRT}, a fourth order was given in \cite{CHP}
and numerical values of the coefficients until 13th order are
contained in \cite{WKO}.

The series (\ref{GapSerN2}) contains a singularity at $\Jp = 0$
which has no physical meaning but is simply due to the choice
of expansion parameter. We therefore analyze it by removing this
singularity via the substitutions
\beq
x = {\Jp \over J+\Jp} \quad ; \qquad
\tilde{x} = \tan^{-1}\left({J \over \Jp}\right) \, .
\label{SingRem}
\eeq
{}From the raw transformed series one can
then find some indication of an extended  massless phase at
small $\Jp$ if $\Delta < \Delta_c$ with $\Delta_c \approx 0.25\ldots0.5$. 
The opening of this massless phase is predicted by the zero-loop
analysis of the previous section to take place at $\Delta_c = 0$.
Since the information obtained in the weak-coupling regime from
a strong-coupling series is not extremely accurate, this agreement
can be considered reasonable.

Now we turn to $N=3$ and OBC. In a way similar to the previously
discussed series one finds the following fourth order series for
the lower and upper boundary of the $\langle M \rangle = 1/3$
plateau:
\bea
{h_{c_1} \over \Jp} &=&
\left(\Delta+1\right) {J \over \Jp}
-{\left (\Delta+1\right )\left (8 \Delta-5\right ) \over 27}
     \left({J \over \Jp}\right)^2
+{\left (\Delta+1\right )\left (142 \Delta^{2}-307 \Delta-23\right )
  \over 972 } \left({J \over \Jp}\right)^3 \nn \\
&& +{\left (\Delta+1\right )\left (40572 \Delta^{3}-83025 \Delta^{2}
          +76961 \Delta-73295\right )
   \over 367416 } \left({J \over \Jp}\right)^4
   + {\cal O}\left(\left({J \over \Jp}\right)^5\right) \, ,
\label{hc1} \\
{h_{c_2} \over \Jp} &=&
{3 \over 2} - {J \over \Jp}
 +{10+17 \Delta^{2} \over 36} \left({J \over \Jp}\right)^2
+{2196 \Delta+252-554 \Delta^{3}+171 \Delta^{2} \over 3888}
     \left({J \over \Jp}\right)^3 \nn \\
&& +{30172+38988 \Delta-28387 \Delta^{2}+7028 \Delta^{3}-8886 \Delta^{4}
  \over 326592} \left({J \over \Jp}\right)^4
   + {\cal O}\left(\left({J \over \Jp}\right)^5\right) \, .
\label{hc2}
\eea
A third-order version of these series was already presented in \cite{CHP}
for the special case $\Delta = 1$. We employ again the transformations
(\ref{SingRem}) to analyze these series. The raw transformed series
indicate for $\Delta = 1$ that the $\langle M \rangle = 1/3$ plateau 
does not extend down until $\Jp = 0$ but ends at a critical value
$\Jpc$. The numerical value is found to be $\Jpc \approx 1.0 \ldots 1.4 J$ at
$\Delta = 1$. This number should however not be taken too seriously
as is also indicated by the large uncertainty of the critical anisotropy
$\Delta_c$ above which this plateau extends over all non-zero $\Jp$:
$\Delta_c \approx 1.0 \ldots 1.6$. At least, this rough estimate
for $\Delta_c$ is compatible with $\Delta_c \approx 1.19$ as obtained from
the zero-loop weak-coupling analysis.

The next case we consider is $N=4$ and PBC. In the strong-coupling
limit we find plateaux at $\langle M \rangle = 0$ and at
$\langle M \rangle = 1/2$. Series can be computed readily for the
gap (which determines the boundary of the $\langle M \rangle = 0$
plateau) and the lower and upper boundary of the $\langle M \rangle = 1/2$
plateau. In this order, they read
\bea
{\Gap_4^{(p)} \over \Jp} &=& 1 - {4 \over 3} \left({J \over \Jp}\right)
   + {33 \Delta^2 - 12 \Delta + 20 \over 108} \left({J \over \Jp}\right)^2
   + {24+194 \Delta^{2}+131 \Delta \over 1296} \left({J \over \Jp}\right)^3
\nn \\
&& + {3524213-17599776 \Delta^{2}+9014208 \Delta+1923768 \Delta^{3}
      +7733988 \Delta^{4} \over 39191040} \left({J \over \Jp}\right)^4
\nn \\
&& + {\cal O}\left(\left({J \over \Jp}\right)^5\right) \, ,
\label{GapSerN4p} \\
{h_{c_1}^{(p)} \over \Jp} &=& 1
   + {3 \Delta + 8 \over 6} \left({J \over \Jp}\right)
   + {9 \Delta^2 + 96 \Delta - 308 \over 864} \left({J \over \Jp}\right)^2
\nn \\
&& + {369 \Delta+972 \Delta^{3}-1314 \Delta^{2}-9464 \over 31104}
               \left({J \over \Jp}\right)^3 \nn \\
&& - {885195 \Delta^{4}-69076728 \Delta^{2}- 61318885-545832 \Delta^{3}
      +117897360 \Delta \over 156764160} \left({J \over \Jp}\right)^4
\nn \\
&& + {\cal O}\left(\left({J \over \Jp}\right)^5\right) \, ,
\label{hc1N4p} \\
{h_{c_2}^{(p)} \over \Jp} &=& 2
   + {\Delta  - 2 \over 2} \left({J \over \Jp}\right)
   + {5 \Delta^2 + 22 \over 32} \left({J \over \Jp}\right)^2
   - {8 \Delta^{3}-42 \Delta-9 \Delta^{2}-51 \over 256}
                                     \left({J \over \Jp}\right)^3 \nn \\
&& + {38 \Delta^{4}-1981 \Delta^{2}-56 \Delta^{3}+1634 \Delta -403 \over
      4096} \left({J \over \Jp}\right)^4
   + {\cal O}\left(\left({J \over \Jp}\right)^5\right) \, .
\label{hc2N4p}
\eea
The superscript `$(p)$' means that these series are for PBC.

Again, we analyze these series using the transformations
(\ref{SingRem}). We apply this first to the gap (\ref{GapSerN4p})
and find that the gap closes for some $\Jp > 0$ if $\Delta < \Delta_c$
where the estimates for the critical value span an interval
$\Delta_c \approx 0.8\ldots1.2$. This interval is centered
around the value $\Delta_c = 1$ predicted by power counting in
the context of Abelian bosonization.

Concerning the opening of the $\langle M \rangle = 1/2$ plateau,
we can first locate its ending-point in the same way as before as
$\Jpc \approx 0.8 \ldots 1.6 J$
at $\Delta = 1$. What is more interesting is the conclusion that
this ending-point cannot be pushed down to $\Jpc = 0$ by increasing
$\Delta$. This is in agreement with the zero-loop weak-coupling analysis
which implies that an $\langle M \rangle = 1/2$ plateau does not
exist for $\Jp \ll J$ and $N=4$ regardless of the choice
of $\Delta$.

Finally we present second order versions of analogous series for $N=4$
and OBC (denoted by a superscript `$o$'):
\bea
{\Gap_4^{(o)} \over \Jp} &=& {1\over 2} \left(1+\sqrt{3}-\sqrt{2} \right)
   - \left({1 \over \sqrt{6}} + {2 \over 3}\right) {J \over \Jp}
- \Biggl\{ {\sqrt{6} \over 1104}
               \left(764 \Delta^2 - 1288 \Delta + 947\right) \nn \\
&& \quad
     - {\sqrt{3} \over 1656} \left(1682 \Delta^2 - 2760 \Delta + 1847\right)
     + {\sqrt{2} \over 3312} \left( 4176 \Delta^2 - 7176 \Delta + 4063\right)
\nn \\
&& \quad
       - {1 \over 414} \left(862 \Delta^2 - 1242 \Delta + 869\right)
    \Biggr\} \left({J \over \Jp}\right)^2
   + {\cal O}\left(\left({J \over \Jp}\right)^3\right) \, ,
\label{GapSerN4o} \\
{h_{c_1}^{(o)} \over \Jp} &=& {1\over 2} \left(1+\sqrt{3}-\sqrt{2} \right)
   +{2 \sqrt{6} + 9 \Delta + 8 \over 12} \left({J \over \Jp}\right)
+ \Biggl\{ {\sqrt{6} \over 1656}
              \left(1146 \Delta^2 - 1932 \Delta - 155\right) \nn \\
&& \quad
     + {\sqrt{3} \over 1656} \left( - 1406 \Delta^2 + 2691 \Delta + 637\right)
     + {\sqrt{2} \over 26496} \left(27819 \Delta^2 - 57408 \Delta - 97216\right)
\nn \\
&& \quad
     + {1 \over 828} \left( - 1425 \Delta^2 + 2553 \Delta + 3920\right)
    \Biggr\} \left({J \over \Jp}\right)^2
   + {\cal O}\left(\left({J \over \Jp}\right)^3\right) \, ,
\label{hc1N4o} \\
{h_{c_2}^{(o)} \over \Jp} &=& 1 + {1 \over \sqrt{2}}
   + {\Delta - 4  \over 4} \left({J \over \Jp}\right)
   + {27 \sqrt{2} \Delta^2 + 416 \sqrt{2} - 528\over 128}
          \left({J \over \Jp}\right)^2
   + {\cal O}\left(\left({J \over \Jp}\right)^3\right) \, .
\label{hc2N4o}
\eea
A second order expansion of the dispersion relation at $\Delta = 1$
has already been presented before \cite{RRT}, though with floating-point
coefficients. Eq.\ (\ref{GapSerN4o}) agrees with the result of \cite{RRT}
for the gap $\omega^{-}(k = \pi)$ up to first order, but there is a minor
difference in the second order: We believe that the coefficient of
$\cos{2 k}$ in eq.\ (23) of \cite{RRT} should read $-0.52781\ldots$
(not $-0.469$). We have also checked eq.\ (24) loc.cit.\ and in
this case found perfect agreement.

Given the low order of the series (\ref{GapSerN4o}--\ref{hc2N4o})
we do not try to draw conclusions for the weak-coupling region
from them. We have restricted to only second order since a symbolic
computation of higher orders is very difficult. This is due to 
the many square roots encountered, as is already indicated
by the results presented here.

\section{The strong-coupling effective Hamiltonian of a frustrated ladder}

Here we look at strong coupling ($\Jp \gg J$) for PBC and odd $N$.
In this case additional degeneracies preclude a simple analysis
as in the preceding section. From a first-order consideration in $J$
one infers that the low-energy effective Hamiltonian for (\ref{hamOp})
with $\Delta = 1$ and $h=0$ is then given by (see \cite{ToSu,Schulz}
for $N=3$ and \cite{KaTa} for larger $N$):
\beq
H^{(N,p)}_{{\rm eff.}} = {J \over N}
\sum_{x=1}^L \left(\id + \alpha_N \left(\sigma_x^{+} \sigma_{x+1}^{-}
 + \sigma_x^{-} \sigma_{x+1}^{+} \right) \right) \Sv_x \Sv_{x+1} \, ,
\label{lowEn3p}
\eeq
where the $\Sv_x$ are $su(2)$ operators acting in the spin-space
and $\sigma_x^{\pm}$ act on another two-dimensional space which
comes from a degeneracy due to the permutational symmetry of the
chains.

We have checked the validity of (\ref{lowEn3p}) for
$N=3$, $5$, $7$ and $9$ in the following way: First one has
to determine the groundstate space at each rung for $J=0$ which
is nothing but the groundstate space of an $N$-site Heisenberg
chain. For $N$ odd, the lowest energy states have $S^z = \pm 1/2$.
In the case of OBC, this would be the only degeneracy. $SU(2)$
symmetry is then sufficient to conclude that the effective
Hamiltonian is a simple Heisenberg chain which is gapless
in accordance with the generalized Haldane conjecture.

For PBC there is another two-fold degeneracy in addition to
this two-fold degeneracy in spin space: For $N$ odd and PBC
the groundstates of a Heisenberg chain carry momenta
$k = \pm 2 \pi \lfloor (N+1)/4\rfloor / N$ where parity symmetry
is reflected in the freedom of choice of sign. So, the groundstate-space
at each rung $x$ is 4-dimensional: The operators $\Sv_x$ act
in the two-dimensional spin space and the $\sigma_x^{\pm}$
act in the two-dimensional space spanned by the groundstate momenta.

This degeneracy makes perturbation expansions in $J$ highly non-trivial:
At first order in $J$ one has to diagonalize the matrix (\ref{lowEn3p})
which is determined by the matrix elements of the interaction
along the legs in (\ref{hamOp}). That the only non-zero
matrix elements are those given in (\ref{lowEn3p}) can be inferred
just from the following symmetries of the full Hamiltonian:
Global $SU(2)$ symmetry (actually one needs only the $U(1)$ Cartan
subalgebra of $su(2)$) and invariance under simultaneous translations
or reflections along all the rungs. These symmetries also imply some
identities between the non-zero matrix elements, but at the end one
still has to explicitly compute some matrix elements -- at least in
order to determine the constants $\alpha_N$. We have performed such
direct computations of matrix elements for $N=3$, $5$, $7$ and $9$
and found the associated values of $\alpha_N$ to be
\beq
\alpha_3 = 1 \, , \qquad
\alpha_5 = {16 \over 9} \, , \qquad
\alpha_7 = 2.6206859\ldots \, , \qquad
\alpha_9 = 3.5012083\ldots \, .
\label{valAlphaN}
\eeq
In contrast e.g.\ to the XXZ-chain (\ref{HamXXZ}), already for $N=3$
the Hamiltonian (\ref{lowEn3p}) does not satisfy the Reshetikhin criterion
(eq.\ (3.20) on p.\ 101 of \cite{Resh}).
Therefore, it is in general not integrable (in the sense that it would
be the Hamiltonian of a one-parameter family of transfer matrices
which commute among themselves and with this Hamiltonian). So, one
has to treat it by other approximate or numerical methods; e.g.\
a DMRG study was carried out for $N=3$ in \cite{KaTa} providing evidence for
a gap to $S=1$ excitations.

In the present paper we are interested in generic XXZ-anisotropies
$\Delta \ne 1$ in the interaction along the chains in (\ref{hamOp})
and thus we should generalize (\ref{lowEn3p}). This generalization
is obvious from the way how the XXZ-anisotropy appears in (\ref{hamOp})
and our derivation of (\ref{lowEn3p}): $\Delta$ just multiplies the
matrix elements of the $\Sz$-$\Sz$ interaction. Therefore, the
effective Hamiltonian for generic $\Delta$ is given by
\beq
H^{(N,p)}_{\Delta} = {J \over N}
\sum_{x=1}^L \left(\id + \alpha_N \left( \sigma_x^{+} \sigma_{x+1}^{-}
 + \sigma_x^{-} \sigma_{x+1}^{+} \right) \right)
\left(\Delta \Sz_x \Sz_{x+1} + {1 \over 2}
\left(S^{+}_x S^{-}_{x+1} + S^{-}_x S^{+}_{x+1} \right)\right) \, ,
\label{lowEn3p1}
\eeq
where the parameters $\alpha_N$ remain those given in (\ref{valAlphaN}).
The generalization (\ref{lowEn3p1})
includes in particular the case $\Delta = 0$, corresponding
to two coupled XY-models. Then (\ie\ for $\Delta = 0$) one can apply a
Jordan-Wigner transformation to (\ref{lowEn3p1}). However, even
in this case
one obtains a four-fermion interaction with the effect that the
problem does not simplify (in contrast to the familiar case
of fermion bilinears). In particular, the determination of
the groundstate of (\ref{lowEn3p1}) for $\Delta = 0$ is far
from being straightforward.

As was pointed out in \cite{CHP}, the effective Hamiltonian (\ref{lowEn3p1})
describes the response of (\ref{hamOp}) to a magnetic field  for
$\abs{\langle M \rangle} \le 1/N$ at strong coupling. For $N=3$
(\ie\ $\alpha_3 = 1$) and $\Delta = 1$ we find by exact diagonalization
of (\ref{lowEn3p}) that the transition to $\langle M \rangle = 1/3$
(full magnetization for the effective Hamiltonian) takes place at
$3 h / J = 4.3146, 4.3121, 4.3108, 4.3100, 4.3096$
for $L=8,10,12,14,16$, respectively. This is in reasonable agreement
with numerical values for
the lower boundary of the $\langle M \rangle = 1/3$ plateau
of (\ref{hamOp}) at $\Jp \gg J$ (compare Fig.\ 4 of \cite{CHP}).

\section{Numerical analysis of the strong-coupling effective Hamiltonian}

To learn more about the spectrum of (\ref{lowEn3p1}),
we have performed numerical diagonalizations mainly for
$N=3$ on finite
systems, as was already done in \cite{KaTa} for $\Delta = 1$.
The Hamiltonian has two conserved quantities: $\Sz$
(for $\Delta = 1$ actually the total spin $S$) and a second similar
quantity related to the first factor in (\ref{lowEn3p1}) which
we denote by $\Sigma^z$. The lowest eigenvalues are located
in the $\Sigma^z =0$ sector. First we look at the gap to
the excitations in the $\Sz = 1$ sector. It turns out that 
one can fit the system-size dependence of this gap nicely by
\footnote{
This appears to be also a good fit to the data presented in
Fig.\ 1 of \cite{KaTa} at least for intermediate system sizes
and would lead to results compatible with the exponential fit
used there. Note also that
convergence is typically faster and clearer for the
periodic boundary conditions used here than
for open boundary conditions (used in \cite{KaTa}).
}
\beq
\Gap_{\Sigma^z = 0, \Sz=1}(L) =
\Gap_{\Sigma^z = 0, \Sz=1}(\infty) + {a \over L} \, .
\label{gapFinSize}
\eeq
Estimates for these parameters based on data for lengths up to $L = 14$
are presented in Table \ref{table1} for some values of $\Delta$ and
$N=3$. The numbers in brackets indicate the $1\sigma$-confidence interval
of the fit for the last given digit. Since this ignores possible
other error sources, the true error may be a little larger.
Our result for $\Delta = 1$
($\Gap_{\Sigma^z = 0, \Sz=1}(\infty) = 0.208(1)\, J$)
agrees within error bounds with that of \cite{KaTa}
($\Gap_{\Sigma^z = 0, \Sz=1}(\infty) = 0.27(7)\, J$).
{}From Table \ref{table1} we conclude that the $\Sz = 1$ excitation
of (\ref{lowEn3p1}) is gapped for all $\Delta \ge 0$ and that
there is nothing special about the case $\Delta = 1$
from this point of view.

One comment is in place regarding the form (\ref{gapFinSize}) since
in a gapped situation the convergence should ultimately be exponential
(or at least of order $L^{-2}$ -- see e.g.\ \cite{Pert} and
references therein).
Here we seem to observe a typical crossover phenomenon, \ie\
the small values of the gap imply a large correlation length
such that our system sizes may be well below the correlation length.
In such a range of system sizes one would indeed expect to observe
finite-size corrections which are typical for massless situations. 
Since the corrections should ultimately become smaller,
this would lead to obtaining systematically too small
values of the gap. With the fit (\ref{gapFinSize}) we thus obtain
a lower bound for the gap which is presumably not far from the true
value. In particular, we can safely infer the presence of a gap. 

\begin{table}[ht]
\begin{tabular}{c|cccccc|c}
$\Delta$ & $0$ & $0.2$ & $0.4$ & $0.8$ & $1.0$ & $1.2$ & $1.0$ \\ \hline
$\Gap_{\Sigma^z = 0, \Sz=1}(\infty)/J$
         & $0.139(5)$ & $0.134(8)$ & $0.166(5)$ & $0.200(1)$
           & $0.208(1)$ & $0.214(2)$ & $0.390(6)$ \\ \hline
$a/J$
         & $2.78(4)$  & $3.35(6)$  & $3.19(4)$  & $3.079(7)$
           & $3.088(7)$ & $3.13(2)$  & $5.72(4)$ \\
\end{tabular}
\caption{
Parameters for the fit (\ref{gapFinSize}) to the $\Sz = 1$ gap of
(\ref{lowEn3p1}). The first six columns are for $N=3$, but various
values of $\Delta$. The rightmost column is for ``$N=\infty$''
with $\alpha_N/N \to 1$.
\label{table1}
}
\end{table}

Concerning the case of $N > 3$, one observes from (\ref{valAlphaN})
and a further value for $\alpha_{11}$ \cite{KaTa}
that $\alpha_N$ is roughly proportional to $N$, \ie\ $\alpha_N
\approx 0.44 \, N$ for large $N$. Using this information,
we have extrapolated (\ref{lowEn3p}) to infinite $N$ setting
$\lim_{N\to \infty} \alpha_N/N = 1$ in order to avoid the uncertainty in the
true proportionality constant. This limit eliminates the
term $\id \times \Sv_x \Sv_{x+1}$ in (\ref{lowEn3p}). The rightmost
column in Table \ref{table1} shows the value for the $S=1$ gap
that we obtain in this case. It should be noted that the properly
rescaled value for $N \to \infty$ is slightly lower than that
for $N = 3$ at $\Delta = 1$ (the former is about $80\%$ of the
latter). However, even for $N=\infty$ our estimate for the gap is
still remarkably distinct from zero. This suggests a gap in the
strong-coupling limit (\ref{lowEn3p}) for all $N$ which slightly
decreases as $N \to \infty$, but does not close even in this limit. 

Now we turn to the ``gap'' in the $\Sz = 0$ sector for $N=3$.
The data in Table \ref{table2} can be interpreted as
evidence that it asymptotically decreases roughly as
\beq
\Gap_{\Sigma^z = 0, \Sz=0} \sim {1 \over L^2} \label{gapS0} \, ;
\eeq
at least this ``gap'' clearly tends to zero in the thermodynamic
limit (in particular close to $\Delta = 0$ the finite-size
exponent could be different from that given in (\ref{gapS0})).
This energy level corresponds to the state constructed in the
generalized Lieb-Schultz-Mattis theorem \cite{LSM,Affleck,Rojo,AOY}.
According to \cite{KaTa} this energy-level should be interpreted
as a degenerate groundstate arising due to spontaneous dimerization
in the thermodynamic limit. The fact that the energy levels 
in Table \ref{table2} have momentum $\pi$ relative to the
groundstate is compatible with this interpretation and yields
$l=2$ for the condition (\ref{condMgen}).

\begin{table}[ht]
\begin{tabular}{c|cccccc}
$\Delta$ & $0$ & $0.2$ & $0.4$ & $0.8$ & $1.0$ & $1.2$ \\ \hline
$L$  & \multicolumn{6}{c}{$L^2 \, \Gap_{\Sigma^z = 0, \Sz=0}/J$} \\ \hline
$4$  & $11.04294$ & $12.16715$ & $13.29556$ & $15.54226$
        & $16.65656$ & $17.76262$ \\
$6$  & $14.71647$ & $15.34371$ & $16.29211$ & $18.64172$
        & $19.94276$ & $21.29515$ \\
$8$  & $17.68885$ & $17.48479$ & $18.05673$ & $20.22660$
        & $21.6$     & $23.0955$ \\
$10$ & $20.18126$ & $18.90011$ & $19.00079$ & $20.85874$
        & $22.23517$ & $23.80763$ \\
$12$ & $22.28842$ & $19.75430$ & $19.34576$ & $20.83216$
        & $22.17023$ & $23.76822$ \\
$14$ & $24.06309$ & $20.15994$ & $19.24047$ & $20.33604$
        & $21.60791$ & $23.19367$ \\
\end{tabular}
\caption{
Rescaled $\Sz = 0$ ``gaps'' of (\ref{lowEn3p1}) with $N=3$
for various values of $\Delta$.
\label{table2}
}
\end{table}

\begin{figure}[ht]
\psfig{figure=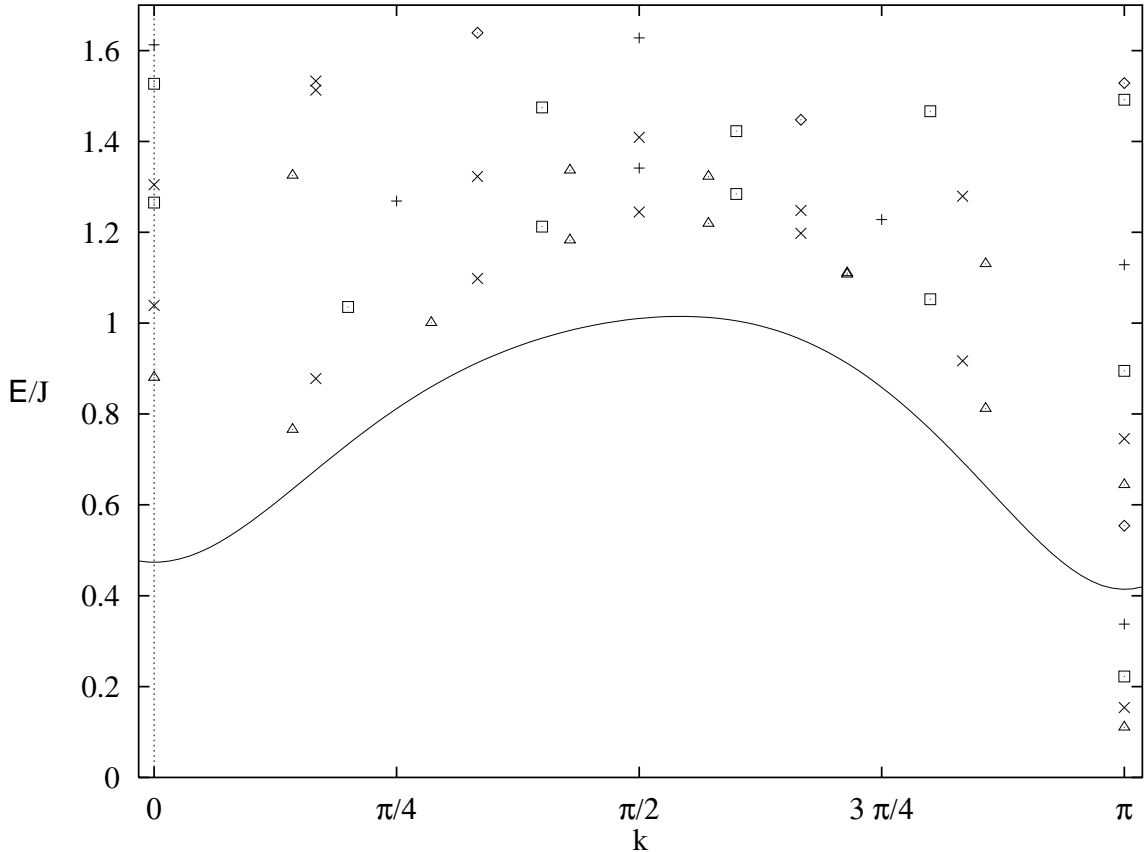,width=\columnwidth,angle=270}
\caption{
Lowest gaps of (\ref{lowEn3p}) with $N=3$ in the sector with
$\Sigma^z = 0$ and total spin $S=0$ as a function of
momentum $k$ relative to the groundstate.
The symbols are for $L=6$ (\SymbolB), $L=8$ ($+$), $L=10$
(\SymbolA), $L=12$ ($\times$) and $L=14$ (\SymbolC), respectively.
The line is the extrapolation (\ref{dispS0}) of the lower boundary
$L \to \infty$.
\label{figure1}
}
\end{figure}

\begin{figure}[ht]
\psfig{figure=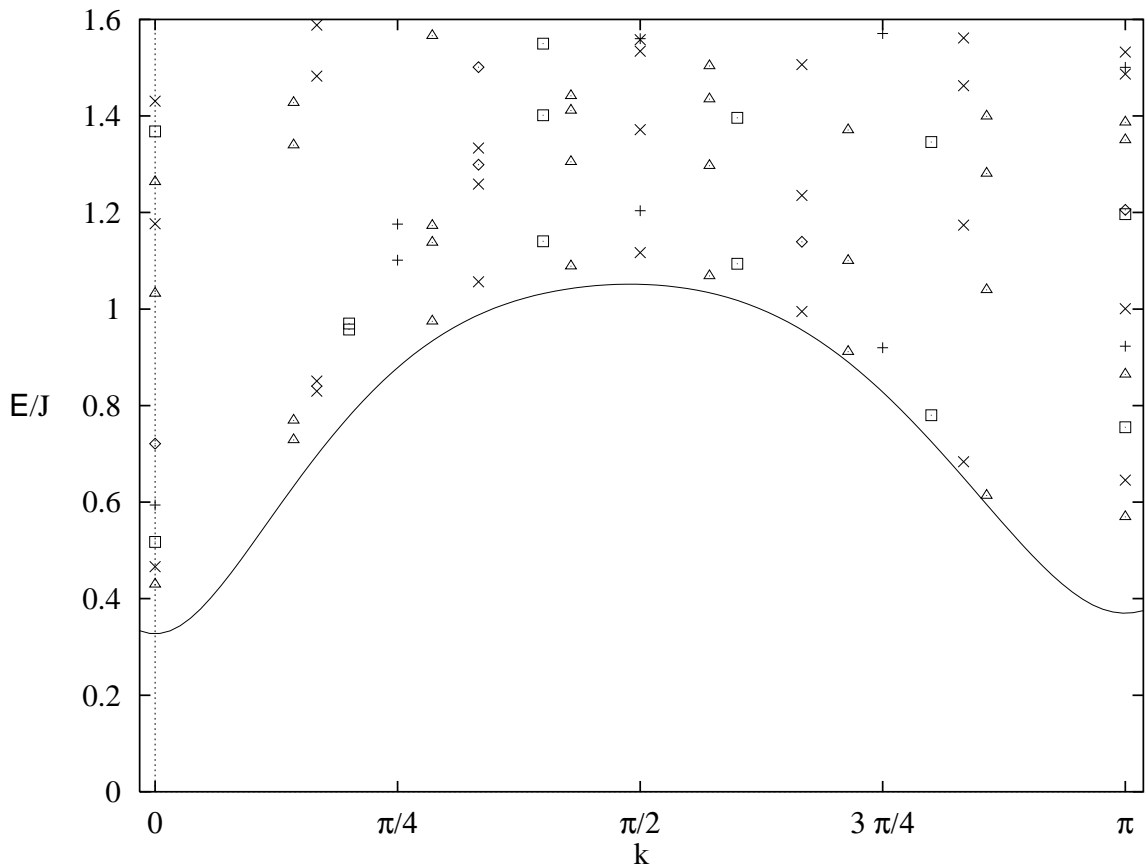,width=\columnwidth,angle=270}
\caption{
Same as Fig.\ \ref{figure1}, but for total spin $S=1$.
The line shows the extrapolation (\ref{dispS1}).
\label{figure2}
}
\end{figure}

Next, we investigate the momentum-dependence of the gaps to the lowest
excited states of (\ref{lowEn3p}) with $N=3$. The data for
$\Sigma^z = 0$ and total spin $S=0$ is shown in Fig.\ \ref{figure1}
and that for total spin $S=1$ (also $\Sigma^z = 0$)
in Fig.\ \ref{figure2} (compare also Fig.\ 4 of \cite{KaTa}).
Here, we measure the momentum of the excitations relative to
the groundstate.
It should be noted that due to parity conservation only half
of the spectrum is shown (the parts for $k > \pi$ or $k < 0$ are
mirror-symmetric extensions of this figure).
The two figures look quite similar. Both can be interpreted
as the lower boundary of a (two-particle) scattering
continuum, In particular, we do not observe one-particle states.

To extrapolate the lower boundaries of these two-particle scattering
states, we have Fourier transformed $\Gap^2$
\footnote{
We Fourier transform $\Gap^2$ rather than $\Gap$, since then
higher harmonics are better suppressed, as must be the case
for the scaling limit of a lattice model close to a second
order critical point.
}.
Then we have extrapolated each coefficient of the Fourier series separately
using a Shanks transform (which is the $\alpha= 0$ special case of
the vanden Broeck--Schwartz algorithm -- see e.g.\ \cite{HeSchue}).
This leads to:
\bea
\Gap^2_{\Sigma^z = 0, S = 0}(k)/J &=&
0.654 (478)
- 0.014 (191) \cos{k}
- 0.411 (108) \cos{2k} \nn \\ &&
+ 0.040 (136) \cos{3k}
- 0.044 \cos{4k} \, , \label{dispS0} \\
\Gap^2_{\Sigma^z = 0, S = 1}(k)/J &=&
0.671 (43)
+ 0.023 (24) \cos{k}
- 0.492 \cos{2k} \nn \\ &&
- 0.025 \cos{3k} 
- 0.058(11) \cos{4k} 
- 0.013(50) \cos{5k} \, . \label{dispS1}
\eea
The numbers in brackets indicate estimates for the error of the
last given digits. Here we have suppressed the highest harmonics,
since they cannot be reasonably extrapolated. but are expected to
be small anyway.

In this way we obtain a rather inaccurate estimate for the gap
$\Gap \approx 0.3 J$ with a large uncertainty which is due to
the large errors in particular in (\ref{dispS0}) and the
uncertainty in the higher harmonics. Nevertheless, this estimate
is still quite close to the one in Table \ref{table1}.
A more interesting observation is that (\ref{dispS0}) and 
(\ref{dispS1}) are equal within error bounds. This suggests that
these two thresholds can be interpreted in terms of two-particle
scattering states of a single fundamental particle. Such a fundamental
excitation would have to be similar to the spinon in the XXZ-chain;
in particular it would have to carry $S = 1/2$ (and $\Sigma^z = \pm 1/2$).

Let us now try to exhibit this fundamental excitation explicitly.
For even $L$ and periodic boundary conditions we have only
found two-particle scattering states in the low-lying excitation
spectrum. Therefore, it is natural to look for a spinon-type
excitation at odd $L$ (still periodic boundary conditions) in
the same way as one can exhibit the spinon for the XXZ-chain
\cite{FaTa,GoSa,RHSZ}. We have
computed the spectrum of (\ref{lowEn3p}) for $N=3$ and odd $L$ from $5$ to
$13$ in the sector with $\Sigma^z = 1/2$ and total spin $S=1/2$. The main
difference between the present situation and the XXZ-chain is
that here we expect a charge conjugate pair of spinons
($\Sigma^z = \pm 1/2$) while for the XXZ-chain there was only
one.

\begin{figure}[ht]
\psfig{figure=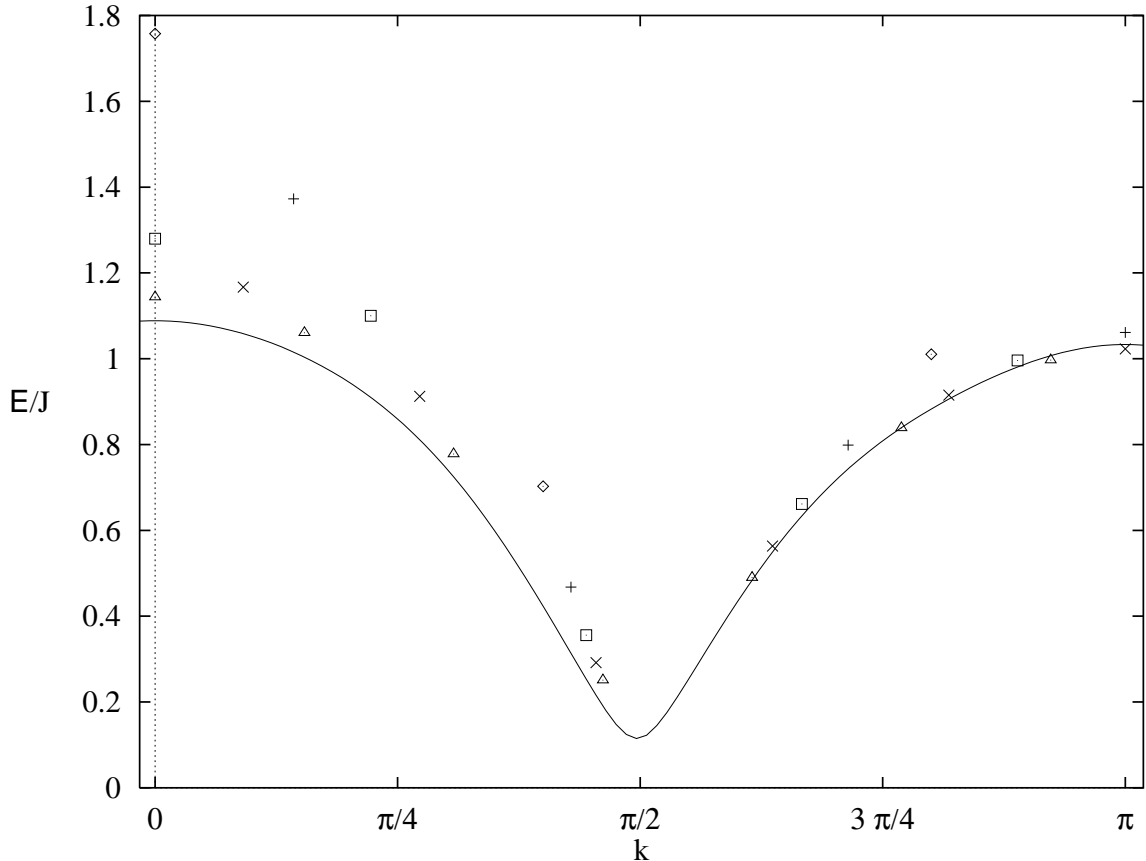,width=\columnwidth,angle=270}
\caption{
The spinon of (\ref{lowEn3p}) with $N=3$, \ie\ lowest gaps in
the sector with $\Sigma^z = 1/2$ and total spin $S=1/2$.
The symbols are for $L=5$ (\SymbolB), $L=7$ ($+$), $L=9$
(\SymbolA), $L=11$ ($\times$) and $L=13$ (\SymbolC), respectively.
The line is the extrapolation (\ref{dispSpinon}) of the dispersion
curve to the thermodynamic limit.
\label{figSpinon}
}
\end{figure}

Making single-particle states visible is traded for the absence of
a groundstate at odd $L$. In order to permit interpretation of the
results as gaps we have therefore
interpolated the groundstate energy using the values at $L \pm 1$.
The resulting dispersion relation for the spinon is shown in
Fig. \ref{figSpinon}. As was already the case for the spectra
at even $L$, it turns out that $k$ should be defined such that
translationally invariant states on the lattice appear alternately
at $k=0$ and $k=\pi$ -- the actual convention can be read off from 
Fig. \ref{figSpinon} noting that only either $k=0$ or $k=\pi$ can be
realized for odd $L$.

To interpret the data, we have again Fourier transformed $\Gap^2$.
Firstly, this gives an interpolation of $\Gap_{\Sigma^z = 1/2, S = 1/2}$
at $k = \pi/2$. Analogously to (\ref{gapFinSize}) we fit the data for
$L=5,9,13$ to the form (the values for $L=7$ and $11$ should be omitted
to obtain a monotonic sequence)
\beq
\Gap_{\Sigma^z = 1/2, S = 1/2}(L) =
\Gap_{\Sigma^z = 1/2, S = 1/2}(\infty) + {\bar{a} \over L} \, ,
\label{gapSpinFinSize}
\eeq
and obtain an estimate for the gap of the spinon
$\Gap_{\Sigma^z = 1/2, S = 1/2}(\infty) = 0.131(8) \, J$ with
$\bar{a} = 0.51(6)\, J$. This is roughly consistent with half
the value in Table \ref{table1} or that given in \cite{KaTa},
as it should be if our interpretation as single- respective two-spinon
scattering states is correct.

An alternate way to analyze the data is to extrapolate each coefficient
of the Fourier series separately using a Shanks transform. Using
now all available $L$, we find
\bea
\Gap^2_{\Sigma^z = 1/2, S = 1/2}(k)/J &=&
0.6331 (155)
+ 0.0592 (184) \cos{k} \nn \\ &&
+ 0.5387 (122) \cos{2k}
+ 0.0121 (63) \cos{3k} \label{dispSpinon} \\ &&
- 0.0633 (160) \cos{4k}
- 0.0127 \cos{5k}
+ 0.0177 \cos{6k} \, . \nn
\eea
As before,
the numbers in brackets indicate estimates for the error of the
last given digits. For the two highest harmonics there is not
sufficient data for an extrapolation, so we just take the $L=13$
estimate without being able to estimate an error.
The extrapolation (\ref{dispSpinon}) is shown by the
line in Fig.\ \ref{figSpinon}. Obviously, finite-size effects
are more important for $k < \pi/2$ than for $k > \pi/2$.
Eq.\ (\ref{dispSpinon}) yields another
estimate for the gap of the spinon
$\Gap_{\Sigma^z = 1/2, S = 1/2}(\infty) \approx 0.116 \, J$.
The error estimate obtained from (\ref{dispSpinon}) is not
sensible, but the value for the gap itself is very close to
our previous extrapolation or half the value given in
Table \ref{table1}.

Finally, we have checked that within error bounds the dispersion
relation (\ref{dispS1}) can be written in terms of (\ref{dispSpinon}) 
as $\Gap_{\Sigma^z = 0, S = 1}(k) = \Gap_{\Sigma^z = 1/2, S = 1/2}(k-k')
+ \Gap_{\Sigma^z = 1/2, S = 1/2}(k')$ with some $k'$. Such a decomposition
must be possible if our particle interpretation is correct,

Methods similar to the ones used in the present section may be useful
also in other cases beyond the present one and the study of
\cite{RHSZ}. One natural such candidate is a direct observation
of a spinon-type excitation in $N=3$ cylindrically coupled chains
at intermediate or small couplings $\Jp$.

\section{Summary of results}

Our results are best summarized in (schematic) magnetic phase
diagrams. For definiteness we consider the $SU(2)$ symmetric
situation $\Delta = 1$, though similar pictures can be drawn
for other values of $\Delta$ as well.

\begin{figure}[htp]
\hfil\psfig{figure=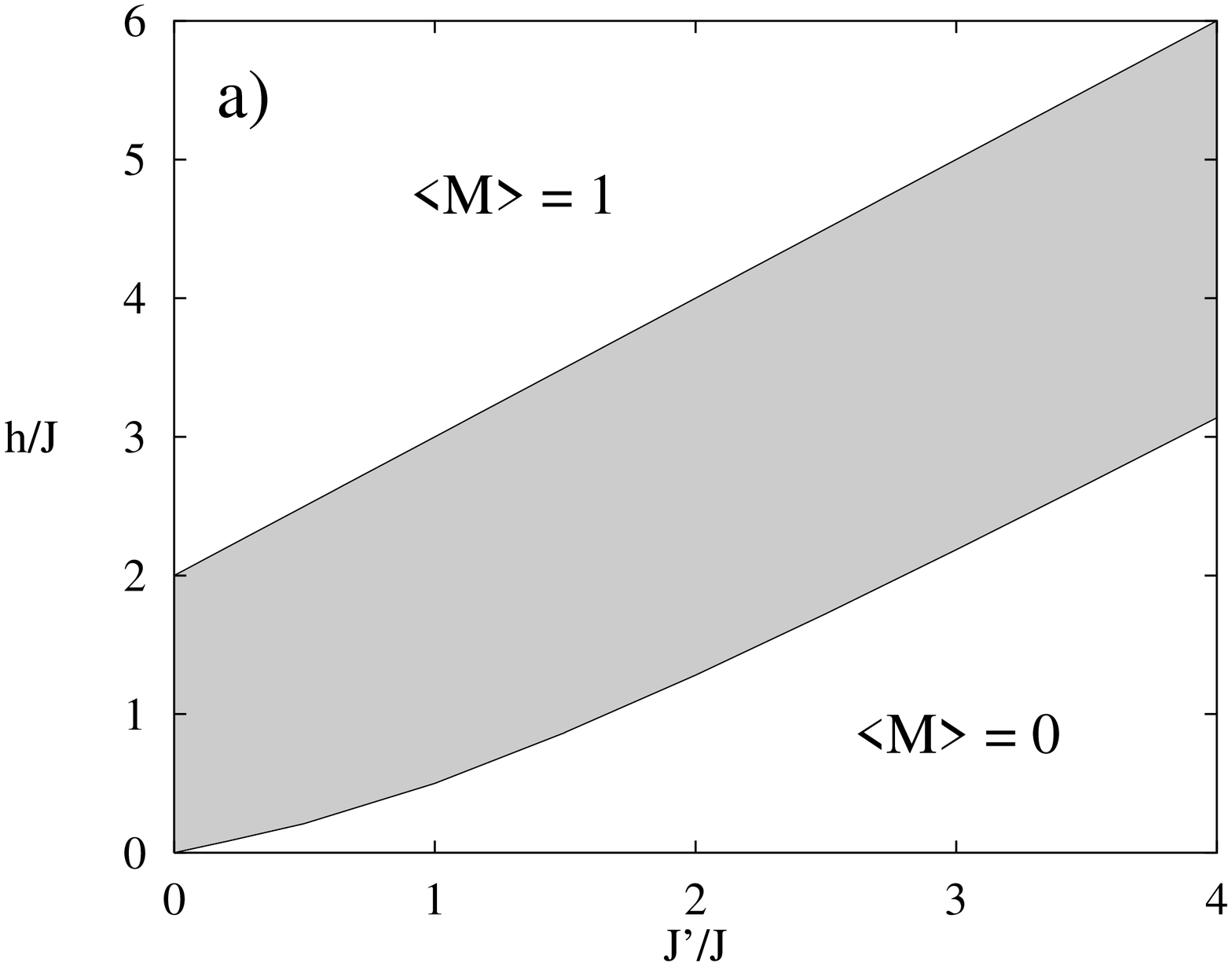,width=0.5\columnwidth,angle=270}\hfil\\
\psfig{figure=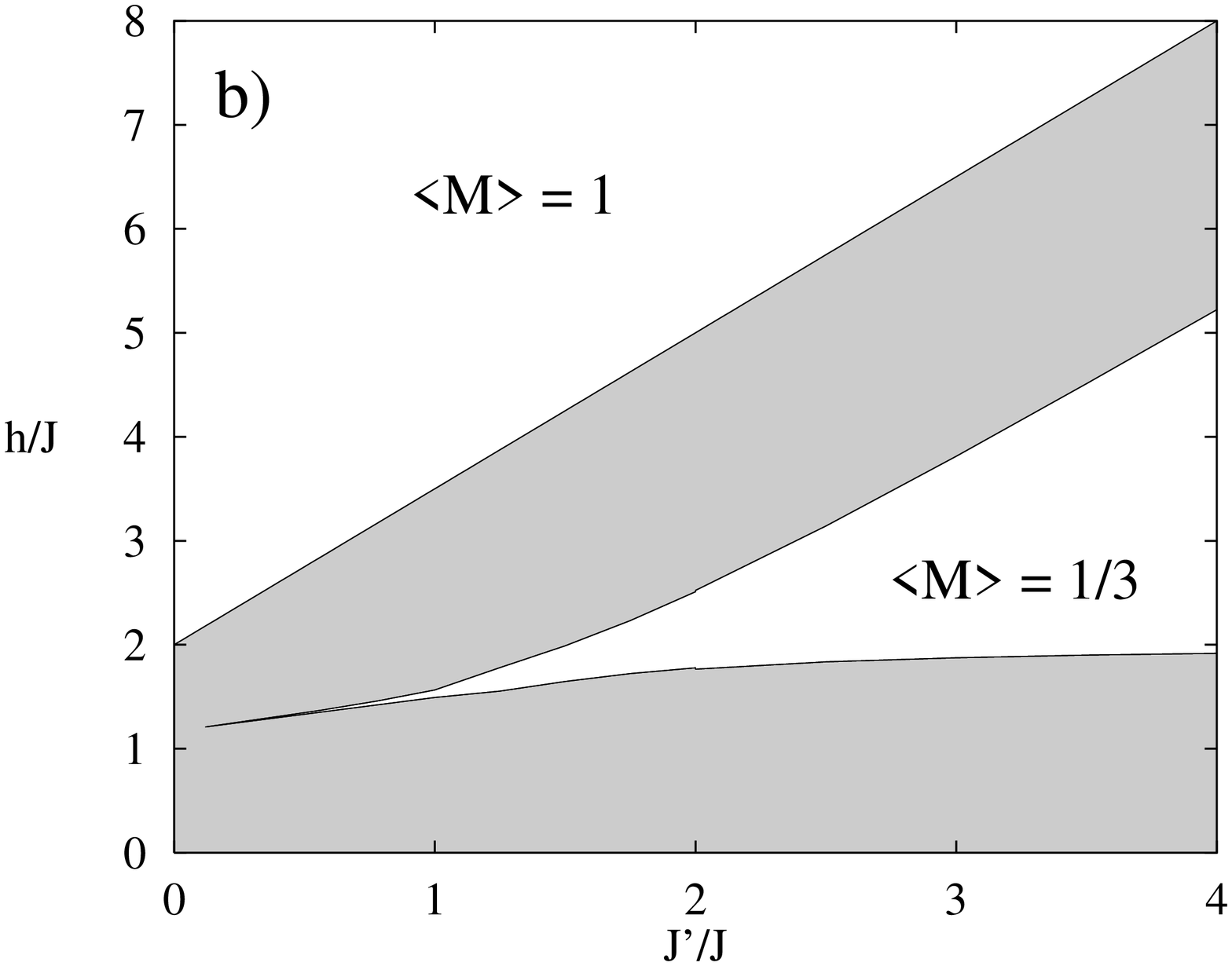,width=0.5\columnwidth,angle=270}\psfig{figure=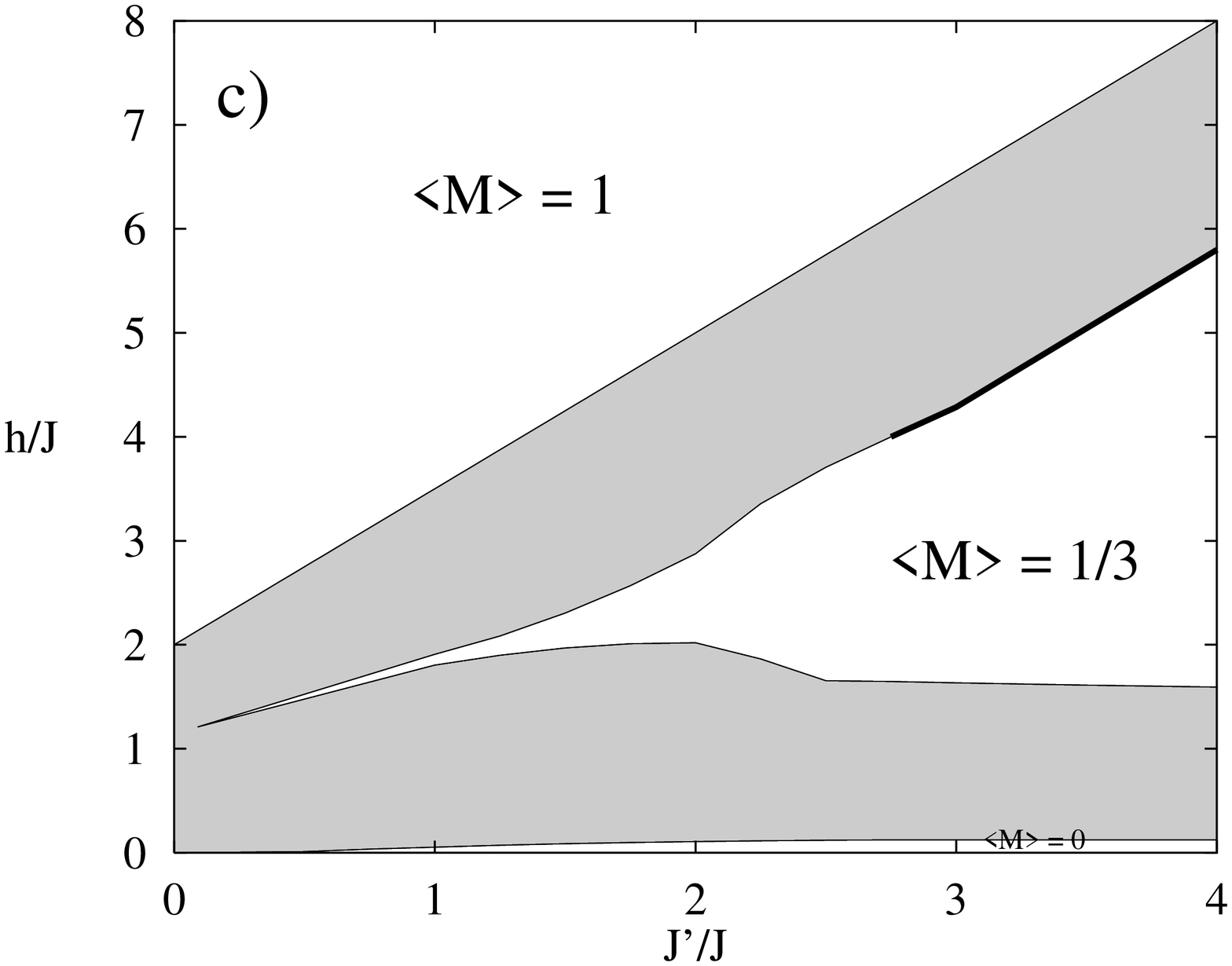,width=0.5\columnwidth,angle=270}
\psfig{figure=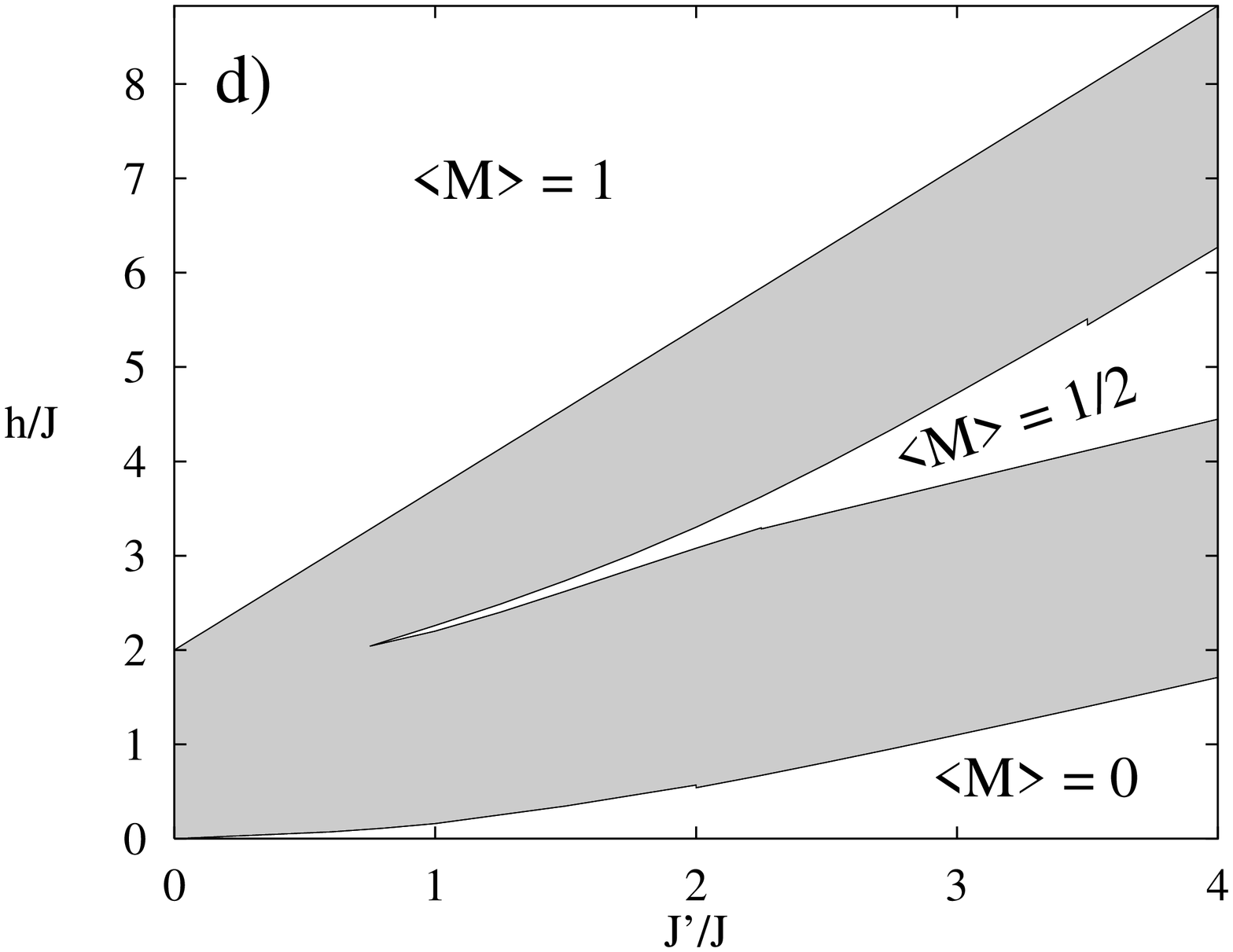,width=0.5\columnwidth,angle=270}\psfig{figure=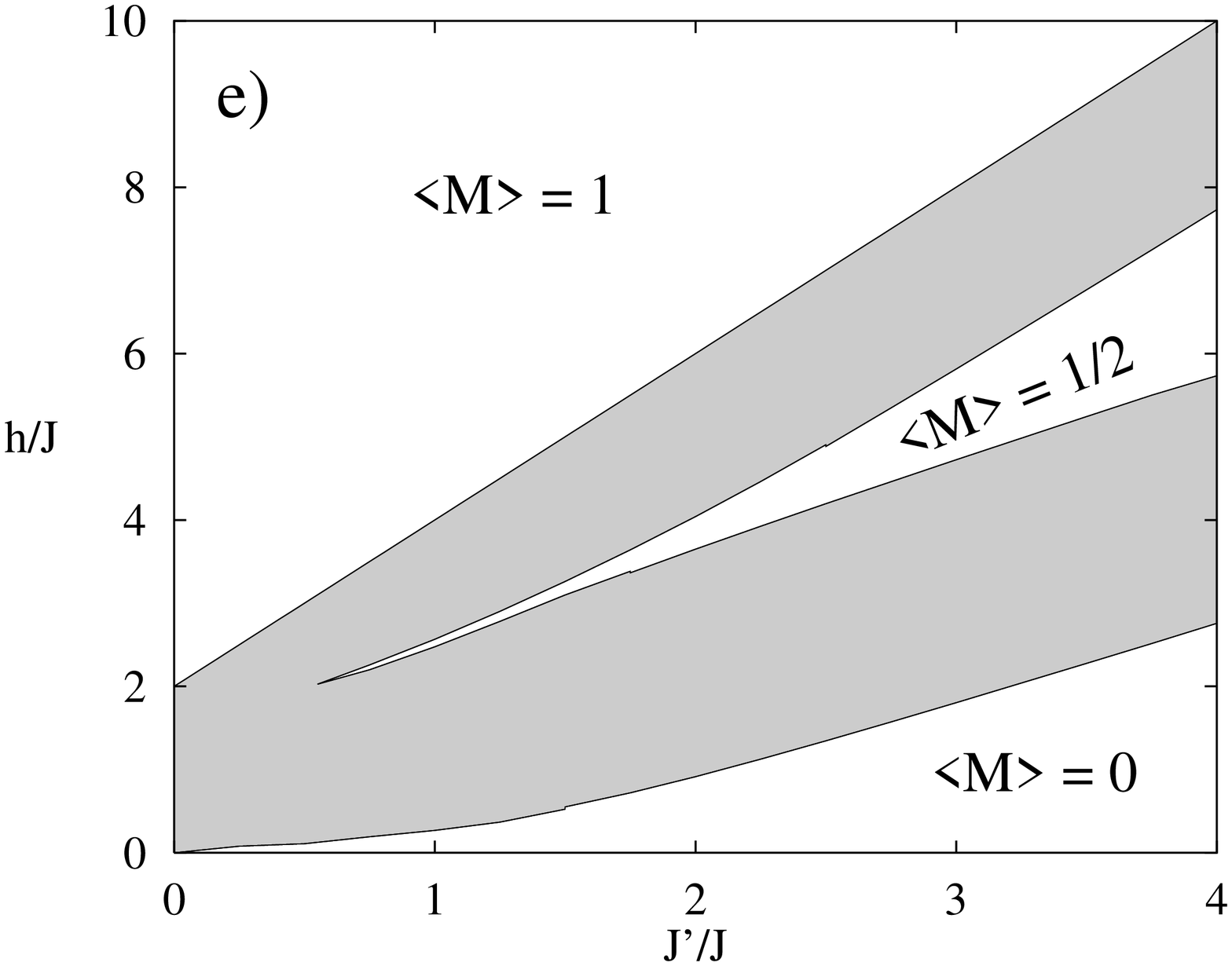,width=0.5\columnwidth,angle=270}
\caption{
Schematic magnetic phase diagram at $\Delta =1$
for a) $N=2$, b) $N=3$ and OBC, c) $N=3$ and PBC,
d) $N=4$ and OBC, e) $N=4$ and PBC.
White regions in the $h$-$\Jp/J$ plane indicate gapped regions with
a plateau in the magnetization curve while the shaded areas
are massless and the magnetization $\langle M \rangle$ changes
continuously if the applied field $h$ is varied.
\label{figMPDs}
}
\end{figure}

For completeness, let us start with the case $N=2$, where
the corresponding picture is given by Fig.\ \ref{figMPDs}a).
The boundary of the $\langle M \rangle = 0$ plateau is determined
by the spin-gap in zero field which for an $N=2$ leg ladder
has been studied in great detail. For $\Jp/J \le 1.5$ we use
the Quantum Monte Carlo results of \cite{BGW} in Fig.\ \ref{figMPDs}a);
for $\Jp/J \ge 1.5$ the raw 13th order strong-coupling series of
\cite{WKO} is used instead (note the excellent matching at $\Jp/J = 1.5$). 

Both the numerical data \cite{BGW} and the series expansions
\cite{WKO} support a linear opening of the gap for small $\Jp$,
as was predicted by a dimensional analysis of the perturbing
operator in the field-theoretic formulation \cite{StroMi,ToSuL,ShNeTs}.

Fig.\ \ref{figMPDs}b) shows a more interesting case, \ie\
$N=3$ with OBC. The boundaries of the $\langle M \rangle = 1/3$
plateau have been determined from the fourth order strong-coupling series
(\ref{hc1}) and (\ref{hc2}) for $\Jp/J \ge 2$. For $1 \le \Jp/J \le 2$
we obtained them from a Shanks extrapolation of the finite-size
data in our earlier paper \cite{CHP}. The remaining weak-coupling
region is the most speculative part of the figure. We located
the ending-point of the $\langle M \rangle = 1/3$ plateau
in the vicinity of the corresponding point on the magnetization
curves of decoupled Heisenberg chains, as is suggested by
the Abelian bosonization analysis if one assumes a similar
behaviour for OBC and PBC (the bosonization analysis predicts for
PBC that the $\langle M \rangle = 1/3$ plateau disappears
for small but non-zero $\Jp/J$).

The analogous case with changed boundary conditions (\ie\
$N=3$ and PBC) is shown in Fig.\ \ref{figMPDs}c). Here, no series
expansions are possible due to extra degeneracies at strong coupling.
The boundaries of the plateaux have therefore been determined
in this case mainly on the basis of older numerical data \cite{CHP}.
We have used a Shanks extrapolation for $L=4$, $6$ and $8$
for $\Jp/J \ge 2.75$ at the lower boundary of the
$\langle M \rangle = 1/3$ plateau and for $\Jp/J = 2.5$ 
at its upper boundary to estimate their location.
For smaller couplings, the finite-size data is non-monotonic. The
best we can do in the range $1 \le \Jp/J \le 2.5$ is to fit the
$L=4$ and $8$ data to a form with $1/L$ corrections like
(\ref{gapFinSize},\ref{gapSpinFinSize}). The ending point
of the plateau is again placed on the basis of the weak-coupling
analysis. It should be noted that the nature of the transition
at the upper boundary of the $\langle M \rangle = 1/3$ plateau
changes qualitatively for $\Jp/J \age 2.75$ \cite{CHP}.
This strongly frustrated region is indicated by the bold line
in Fig.\ \ref{figMPDs}c). It is possible that the
transition becomes first order along this line. Note that
the region in question is far outside the weak-coupling region,
where we expect all transitions to be continuous.

Another interesting difference between Fig.\ \ref{figMPDs}b) and
Fig.\ \ref{figMPDs}c) is that in the latter a tiny
$\langle M \rangle = 0$ plateau (\ie\ a gap) opens.
Its boundary has been estimated at intermediate couplings
by fitting the $L=4$, $6$ and $8$ data \cite{CHP} to the form
(\ref{gapFinSize}). This yields slightly smaller values than
those given in \cite{KaTa} in the cases where we overlap. However,
we agree with \cite{KaTa} in the most important point, namely
the existence of such a plateau. A numerical determination of its
ending point is difficult, and the field-theoretical weak-coupling
analysis is not yet conclusive either \cite{Arrigoni,ToSu} --
the ending point may well be anywhere in the weak-coupling region
$0 \le \Jp \le J$.

Finally, the magnetic phase diagrams for $N=4$ are given by
Figs.\ \ref{figMPDs}d) and e) for OBC and PBC, respectively.
To obtain them, we have performed further numerical computations.
For the upper boundary of the $\langle M \rangle = 1/2$ plateau
we have numerical data for $L=4$, $6$ and $8$ such that we can
apply a Shanks transform to it. For its lower boundary
we have only $L=4$ and $6$ data and therefore we have to make an
assumption on the finite-size corrections to extrapolate it
(we assumed $1/L$ corrections, though this is not entirely
satisfactory). In the strong-coupling region we used our series
instead of numerical data. The series and numerical data are
matched at $\Jp/J = 3.5$ or $\Jp/J = 2.5$ for the series
corresponding to the upper boundary for OBC (\ref{hc2N4o}) or
PBC (\ref{hc2N4p}), respectively. At the lower boundary of
the $\langle M \rangle = 1/2$ plateau we matched the series
(\ref{hc1N4o}) and (\ref{hc1N4p}) to the numerical data at
$\Jp/J = 2.25$ and $\Jp/J = 1.75$, respectively. Neither of the
methods accessible to us is very accurate in the region where
this plateau closes, but all three methods (numerical, series
and Abelian bosonization) point to a location of the ending
point in the region where $\Jp$ and $J$ are of the same order.

The gaps for $N=4$ are taken from our series (\ref{GapSerN4o})
for $\Jp/J \ge 2$ for OBC and (\ref{GapSerN4p}) for $\Jp/J \ge 1.5$
for PBC. For OBC the accurate numerical data for the gap of
\cite{BGW} is used in the weak-coupling region. The corresponding
line in Fig.\ \ref{figMPDs}e) is in comparison rather an educated
guess which is inspired though by $L=4$ and $L=6$ numerical data.
Regarding the series both for the gap and the boundaries
of the $\langle M \rangle = 1/2$ plateau, we observe a trend that
those for PBC can be used for somewhat smaller values of $\Jp/J$
than those for OBC. This is expected since the former are fourth
order but the latter only second order.

Although Fig.\ \ref{figMPDs} is for the particular choice
$\Delta = 1$ there is nothing particular about this case
(at least for non-zero magnetizations), and one would obtain
similar figures for other values of $\Delta$ as well. Further
plateaux may open for $\Delta > 1$. In particular, there should
always be an $\langle M \rangle = 0$ plateau in $N$ coupled
XXZ-chains with $\Delta > 1$, since each such chain is massive
and this should be preserved
at least for sufficiently weak coupling. In the Ising limit
$\Delta \to \infty$ and for non-frustrating boundary conditions
it is easy to see that this is accompanied by breaking of translational
symmetry to a period $l=2$ in the groundstate. In the general case
$\Delta > 1$, such a period $l=2$ reconciles the appearance of
a gap for both even and odd $N$ with (\ref{condMgen}).

The Abelian bosonization analysis predicts all the massless shaded
regions in Fig.\ \ref{figMPDs} to be $c=1$ theories (with
the exception $\Jp = 0$ where one trivially has a $c=N$ theory).
In these regions the exponents governing the asymptotics of the
correlation functions depend continuously on the parameters. Predictions
can be made, however, for the transitions at the boundaries between
such massless phases and plateau regions. The opening of a plateau
when varying $\Jp$ is a transition of K-T type \cite{KoTh}.
Like in the case of the transition at $\Delta = 1$ in the XXZ-chain,
this implies a very narrow plateau after the transition (c.f.\
(\ref{hcXXZcl1})) which makes it difficult to observe numerically
\cite{CHP}. At the transition point the asymptotics of the
correlation functions is governed by the exponents (\ref{ExpKT})
while along the boundaries of the plateaux one has the universal
exponents (\ref{ExpBD}). It should be noted that an attempt to
verify the latter exponents numerically or experimentally is
likely to rather lead to the exponents characteristic for the
transition point if one is sufficiently close to it.

The field-theoretical analysis also predicts
the asymptotic behaviour of the magnetization in a massless
phase but close to a plateau boundary to be given by
the universal DN-PT behaviour \cite{DzNe,PoTa} (\ref{DNPTexp}).
We have in fact numerically verified such a square-root behaviour
close to saturation ($\langle M \rangle \to 1$) at some values
of $\Jp/J$ for $N=2$, $3$ and $4$ with both OBC and PBC. However,
the example of the XXZ-chain shows (c.f.\ Fig.\ \ref{figMagD2})
that close to other plateau boundaries this universal behaviour
may be restricted to a tiny region and its observation could be
very difficult. In experimental situations, it will be further
obscured by thermal fluctuations and other effects such as disorder
(see e.g.\ \cite{HLP}). This explains why rather accurate experiments
on $N=2$ leg spin-ladder materials \cite{CCLMMP,STTKTMG} show no
evidence of a square-root behaviour for $\langle M \rangle \to 0$. 

Quite surprisingly, massless excitations (though non-magnetic
ones) also arise in plateau regions. This can be seen from
(\ref{lowEn3p1}) which for $\langle M \rangle = 1/N$ is just
an XY-chain and therefore massless. This yields massless
excitations in the limit $\Jp \to \infty$ in Fig.\ \ref{figMPDs}c),
or more generally in the strong-coupling limit on the
$\langle M \rangle = 1/N$ plateau for $N$ odd and PBC. Whether
$J/\Jp = 0$ is just a critical point or if massless non-magnetic
excitations also arise at finite $\Jp$ remains to be
investigated.

\section{Discussion and Conclusion}

In this paper we have investigated the conditions under which plateaux
appear in $N$-leg spin ladders as well as the universality classes
of the transitions at the boundaries of such plateaux. 
Certain small plateaux may have slipped our attention. For example,
there could be a narrow $\langle M \rangle = 2/3$ plateau for $N=3$
and PBC at intermediate or strong coupling $\Jp$ which would be accompanied
by spontaneous breaking of translational symmetry to a period $l=2$.
If this should
turn out to be the case, it would have to be added to Fig.\ \ref{figMPDs}c).
However, our main point is the presence of such plateaux,
not the absence of particular ones.

We also confirmed the conclusion of \cite{KaTa} that in the case
$N=3$ frustration induces a zero-field gap at least for sufficiently
strong coupling. It may be even more intriguing that, according
to our strong-coupling data,
this gap seems to survive the $N \to \infty$ limit for an odd number
$N$ of cylindrically coupled chains. This shows that it is
necessary to specify at least boundary conditions along the
rungs in the generalization of the Haldane conjecture to spin
ladders. The peculiar behaviour of a cylindrical configuration may
be interpreted as follows: Strongly frustrating boundary conditions
force a one-dimensional domain wall into the two-dimensional system
corresponding to $N = \infty$. The $N \to \infty$ limit of
the one-dimensional Hamiltonian (\ref{lowEn3p1}) is just the
effective Hamiltonian for the low-energy excitations of this
domain wall. As a consequence, there cannot be any long-range
order which is typical for the two-dimensional Heisenberg model,
and there is no reason why the low-energy spectrum should not
be gapped as it apparently is.

Similar surprises cannot be completely ruled out in the weak-coupling
region for $N$ even and $\langle M \rangle = 0$. The zero-field case is
difficult to control since there is an additional relevant
interaction between the massive degrees of freedom and the possibly
massless ones (the coefficient of $\lambda_2$ in (\ref{LeH})).
In general, this gives rise to non-perturbative renormalization.
In the isotropic case $\Delta = 1$
one can use the $SU(2)$ symmetry \cite{Schulz} to infer the
renormalized radius of compactification of the remaining massless
field. This then leads to the generalized Haldane conjecture
in the framework of Abelian bosonization. For $\Delta > 1$ a gap
is always expected in the weak-coupling regime since already
the decoupled chains are massive. The situation is far less
clear for anisotropies $\Delta < 1$ since then it is not known
how to control the renormalization group flow. An investigation
of this region by other non-perturbative methods would be
interesting. It may also be desirable to perform further checks
of the absence of an extended massless phase in the $SU(2)$ symmetric
situation for even $N \ge 6$.

Beyond a more detailed understanding of the $N$-leg spin-ladder
model (\ref{hamOp}) treated here, a similar investigation of other models
could be interesting. One natural step would be to include
charge degrees of freedom, and see if interesting effects
arise from the interplay of a magnetic field with transport
properties.

Last but not least, it would be desirable to have an experimental
verification of our predictions. We are confident that this is
possible in principle and hope that it will in fact be carried out.

\bigskip

{\it Acknowledgments:}
We would like to thank F.C.\ Alcaraz, G.\ Albertini, F.H.L.\ E{\ss}ler,
T.\ Giamarchi, M.\ Oshikawa, K.\ Totsuka and B.\ Wehefritz for helpful
discussions concerning in particular a single XXZ-chain.
D.C.C.\ is grateful to CONICET and Fundaci\'on Antorchas for
financial support.
We are indebted to the Max-Planck-Institut f\"ur Mathematik
Bonn-Beuel where the more complicated numerical computations 
have been performed.

\end{document}